# Gas Barrier Performance of Graphene/Polymer Nanocomposites

Yanbin Cui [a], S. I. Kundalwal [b] and S. Kumar [a, c *]

[a] Institute Center for Microsystems (iMicro), Department of Mechanical and Materials Engineering (MME), Masdar Institute of Science and Technology, P.O. Box 54224, Abu Dhabi, U.A.E.

[b] Department of Mechanical and Industrial Engineering, University of Toronto, P.O. Box M5S 3G8, Toronto, Canada

[c] Department of Mechanical Engineering, Massachusetts Institute of Technology, Cambridge, MA 02139-4307, United States

[*] Corresponding author: kshanmugam@masdar.ac.ae; skumaar@mit.edu; s.kumar@eng.oxon.org

**Abstract:**

Due to its exceptionally outstanding electrical, mechanical and thermal properties, graphene is being explored for a wide array of applications and has attracted enormous academic and industrial interest. Graphene and its derivatives have also been considered as promising nanoscale fillers in gas barrier application of polymer nanocomposites (PNCs). In this review, recent research and development of the utilization of graphene and its derivatives in the fabrication of nanocomposites with different polymer matrices for barrier application are explored. Most synthesis methods of graphene-based PNCs are covered, including solution and melt mixing, *in situ* polymerization and layer-by-layer process. Graphene layers in polymer matrix are able to produce a tortuous path which works as a barrier structure for gases. A high tortuosity leads to higher barrier properties and lower permeability of PNCs. The influence of the intrinsic properties of these fillers (graphene and its derivatives) and their state of dispersion in polymer matrix on the gas barrier properties of graphene/PNCs are discussed. Analytical modeling aspects of barrier performance of graphene/PNCs are also reviewed in detail. We also discuss and address some of the work on mixed matrix membranes for gas separation.

***Revised manuscript***

*Abbreviations*: phr, weight parts per 100 weight parts polymer; RH, relative humidity; AFG, amine functionalized graphene; BPEI, branched poly(ethylenimine); CVD, chemical vapour deposition; DA-G, dodecyl amine-modified graphene; DA-GO, dodecyl amine functionalized graphene oxide; DA-RGO, dodecyl amine functionalized reduced graphene oxide; DMAc, dimethylacetamide; EFG, exfoliated graphite; EP, epoxy resin; EVOH, ethylene-vinyl alcohol; FG, functionalized graphene; fGO, functionalized graphene oxide; FGS, functionalized graphite sheets; GNPs, graphite nanoplatelets; GONS, graphene oxide nanosheet; HDPE, high-density polyethylene; iGO, isocyanate treated graphite oxide; IIR, poly(isobutylene-isoprene) rubber; LbL, layer-by-layer; LLDPE, linear low density polyethylene; EMIMAc, 1-ethyl-3-methylimidazolium acetate; OTR, oxygen transmission rates; PET, poly(ethylene terephthalate); PLA, polylactide or poly(lactic acid); PAN, polyacrylonitrile; PANI, polyaniline; PC, polycarbonate; PDDA, poly(diallyldimethylammonium) chloride; PE, polyethylene; PEI, polyethylenimine; PEN, poly(ethylene-2,6-naphthalate); PI, polyimide; PMMA, poly(methyl methacrylate); PND, polynorbornene dicarboximide; PP, polypropylene; PPC, poly(propylene carbonate); PS, polystyrene; PU, polyurethane; PVA, poly(vinyl alcohol); PVC, poly(vinyl chloride); RGO, reduced graphene oxide; SEM, scanning electron microscopy; SBR, styrene butadiene rubber; SPVDF, sulfonated polyvinylidene fluoride; TEM, transmission electron microscopy; TRG, thermally reduced graphene; TPU, thermoplastic polyurethane; WAXD, wide-angle X-ray diffraction; XRD, X-ray diffraction; XNBR, carboxylated acrylonitrile butadiene rubber.

---

## 1. Introduction

Due to their functionality, lightweight, ease of processing and low cost, polymers have replaced conventional materials in packaging applications over the past twenty years. Most frequently used polymers in food packaging are polyethylene (PE), polypropylene (PP), polystyrene (PS), polyvinyl chloride (PVC), and polyethylene terephthalate (PET). [1, 2] However, despite their enormous versatility, a limiting property of polymeric materials in food packaging is their inherent permeability to gases and vapors, including oxygen, carbon dioxide, and organic vapors, etc. [2] The penetration of gas into polymer films has a critical effect on their service performance. The barrier properties of polymers can be significantly enhanced by inclusion of impermeable lamellar fillers, such as clay and graphene, with sufficient aspect ratio to alter the diffusion path of gas-penetrant molecules. [3] These nanofillers make the diffusing molecules follow longer and more tortuous pathways to pass through the nanocomposite film (see Figure 1). [4]

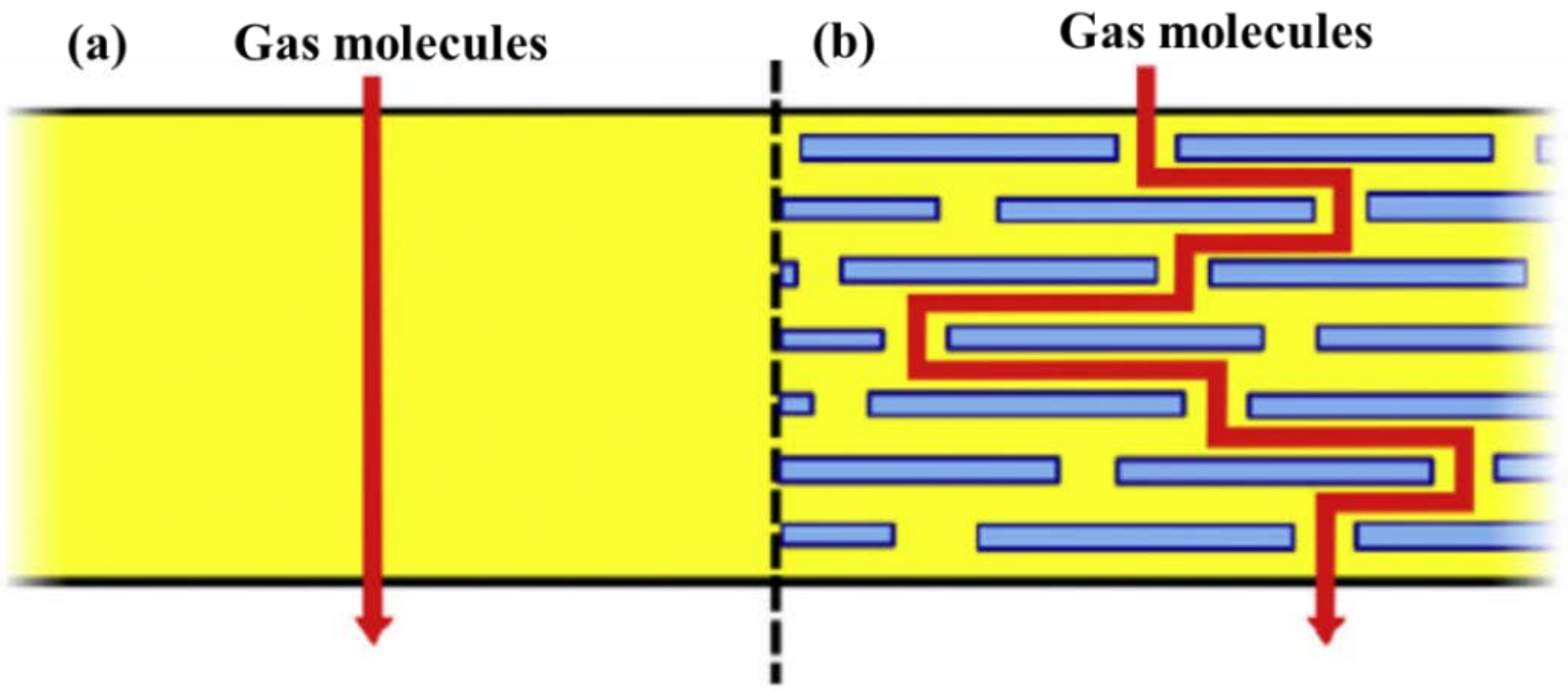

**Figure 1**. Illustration of the "tortuous pathway" created by incorporation of exfoliated nanoplatelets into a polymer matrix film. (a) In a polymer film, diffusing gas molecules migrate via a pathway that is perpendicular to the film orientation. (b) In nanocomposite, diffusing molecules must navigate around impenetrable platelets and through interfacial zones which have different permeability characteristics than those of the pure polymer. [5]

As a result, a significant decline in permeability is observed. [6] The gas barrier performance of PNCs is determined by mainly three factors: filler properties (resistance to gas diffusion, aspect ratio and volume fraction), the intrinsic barrier property of the polymer matrix, and the 'quality' of dispersion (agglomeration/specific interface, free volume generated by mediocre interface management, and the texture/orientation of filler platelets). [7] The crux of successful development of PNCs is coupled with the levels of exfoliation of the layered nanofillers in polymer matrix. [8]

Among the nanofillers, clays are widely used for the barrier applications of PNCs because of their high aspect ratio and their compatibility with various polymers. [9-12] However, hydrophilic clay layers tend to aggregate easily because of their high face-to-face interaction stability due to van der Waals forces. As a result, such aggregation reduces the barrier properties of the resulting PNCs. [4] Compared with clays, graphene-incorporated polymers show not only enhanced gas-barrier properties but also improved mechanical strength, electrical conductivity and thermal properties when properly dispersed in a polymer matrix. [4] For example, graphene has a high mechanical strength (Young`s modulus ~1,100 GPa and fracture strength 125 GPa), [13] thermal conductivity (~5,000 W $m^{-1}$ $K^{-1}$), [14] and electrical conductivity (6000 S/cm) [15] with a high transparency. In the reviewed literature, graphene has been highlighted as another strong candidate for gas-barrier materials because perfect graphene layers do not allow diffusion of small gases through its plane. [4, 16, 17] A large number of studies have been dedicated to investigate the graphene's potential for improving the mechanical, thermal and gas barrier properties of polymers. [18-34]

Over the last few years, graphene has attracted enormous attention due to its unique properties. Recently, Berry has presented a comprehensive review on permeability in graphene and its application in fluid-encasement for wet electron-microscopy, selective gas-permeation, nanopore-bio-diffusion, and barrier coating against rusting and environmental hazards. [35] Nevertheless, there is an emerging trend in using PNCs made up of graphene and its derivatives for gas barrier applications. Derivatives of graphene are graphite oxide and graphene oxide; the latter is produced by exfoliating and reducing the former. [36] Therefore, this review focusses on gas barrier performance of PNCs composed of graphene and its derivatives with a strong emphasis on their formulation methods. A comprehensive review on modeling aspects of gas barrier performance of these PNCs are also presented. We first review the current progress on the production of graphene/PNCs. Then, the gas permeability of PNCs in relation to their structure and processing methods are discussed. Finally, we present theoretical modeling techniques to determine the barrier performance of PNCs.

## 2. Graphene-based polymer nanocomposites

Graphene, a monolayer of $sp^2$-hybridized carbon atoms bonded in the hexagonal lattice, has attracted a vast amount of research interest in recent years. [13, 15, 16, 37-57] It has been viewed as the basic structural unit of all other graphitic carbon allotropes, including graphite, carbon nanotubes and fullerenes (see Figure 2). [55] Because of its superior material properties, graphene has gained significant attention and has become one of the most widely investigated materials. Defect-free, single-crystalline, monolayer graphene has not only excellent mechanical properties [13] and high transparency [44] but is also gas-impermeable. [4, 45, 49] There have been many attempts to use graphene and its derivatives as inorganic nanofillers to enhance the physical properties of PNCs and also to provide enhanced gas barrier properties because of their intrinsic unique properties and good dispersion in common solvents. [4, 48, 58, 59] The relatively high aspect ratio of graphene-based two-dimensional materials can certainly allow much longer pathways for gas-penetrant molecules compared to other nanofillers if they can be fully exfoliated and well-dispersed in PNCs. [60] For example, Compton *et al.* reported that the barrier properties of graphene nanosheets are ~25-130 times higher than that of clay nanofillers at low concentrations. [61, 62] Fully exfoliated and well-dispersed graphene structures in polymer matrix may maximize the gas barrier properties of graphene/PNCs. [4, 60]

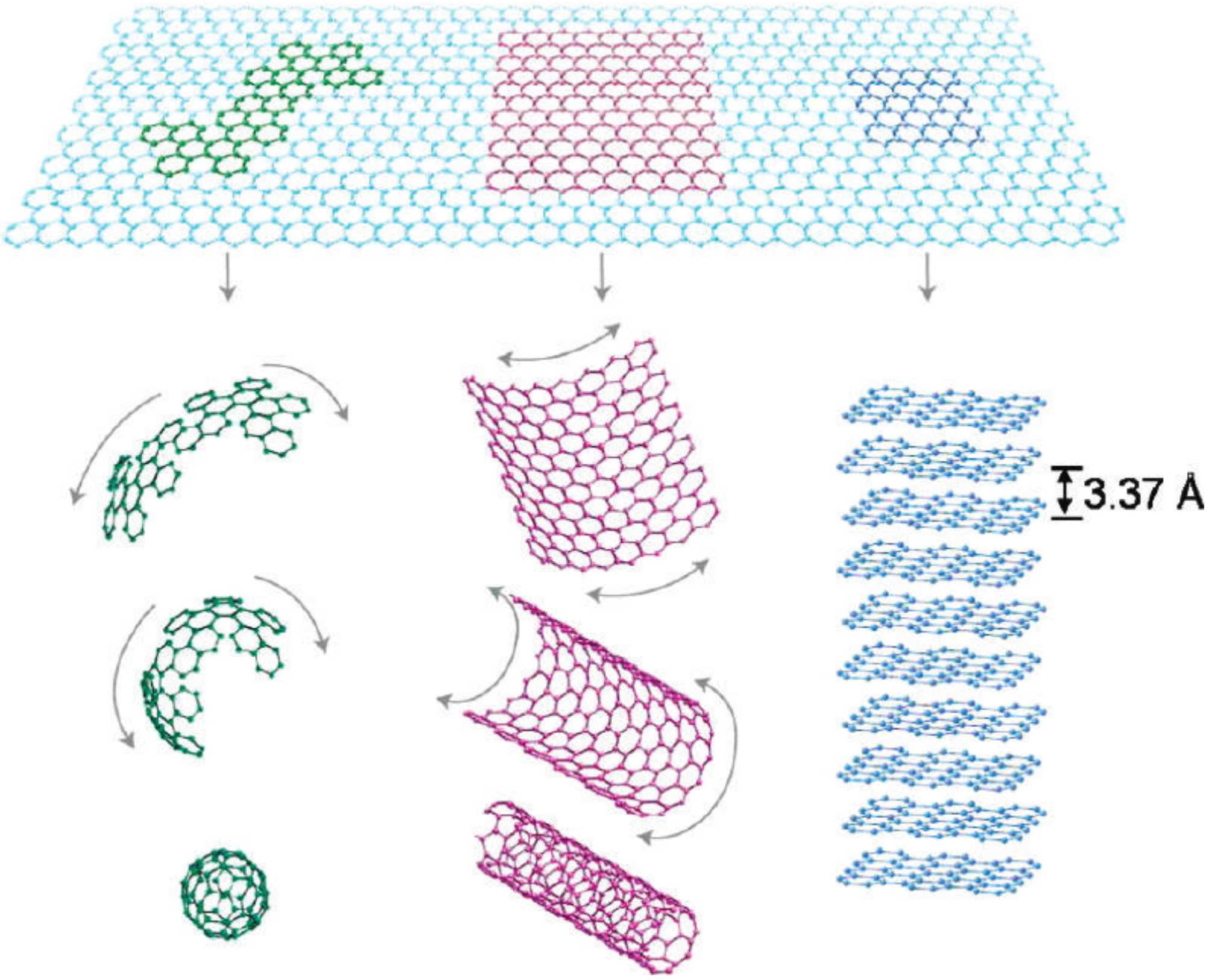

**Figure 2.** Graphene is the building block of all graphitic forms. It can be wrapped up into 0 dimensional buckyballs, rolled into 1 dimensional nanotubes or stacked into 3 dimensional graphite. [58]

## 2.1 Production of graphene and its derivatives

In the past decade, many methods have been reported to prepare graphene of various dimensions, shapes and quality, including epitaxial growth on single-crystal SiC, [42] direct growth on single-crystal metal film [43, 63, 64] or polycrystalline film[54, 65, 66] through chemical vapor deposition (CVD), and chemical reduction of exfoliated graphene oxide layers (see Figure 3). [67-70] The quality and hence the application of graphene are strongly affected by the production method. [56] Although the micromechanical exfoliation, CVD, and epitaxial growth of graphene are suitable for producing high-quality graphene, they have some technical disadvantages which limit the mass-production of graphene. [45] For example, the CVD process is expensive owing to large energy consumption and the underlying metal layer has to be removed. The major drawbacks of synthesis of graphene on SiC are the high cost of the SiC wafers and the high temperatures (above 1,000 $^{o}$C) used. [56] Moreover, CVD and epitaxial growth often produce tiny amounts of large-size, defect-free graphene sheets, which are not a suitable source for PNCs that require a large amount of graphene sheets preferably with modified surface structure. [58] In contrast, the liquid-phase exfoliation of graphite is easy to scale-up in bulk. The liquid-phase exfoliation process is based on exposing the materials to a solvent with a surface tension that favours an increase in the total area of graphite crystallites. With the aid of sonication, graphite splits into individual graphene layers. [52, 56, 71] However, separation of the exfoliated graphene sheets from the bulk graphite could be a challenge. In general, the liquid-phase exfoliation method is suitable for large scale production required for polymer composite applications. Currently, the most promising approaches for large scale production of graphene-based materials are based on exfoliation and reduction of graphite oxide. [37, 58] Generally, graphite oxide is produced using the Hummers methods in which graphite is oxidized using strong oxidants such as $KMnO_4$, $KClO_3$, and $NaNO_2$ in the presence of nitric acid or its mixture with sulfuric acid. [68] Then, graphite oxide is reduced to graphene oxide via chemical or thermal reduction methods. An exfoliated graphene oxide sheets are produced by chemical reduction method. And a stable colloids of graphene oxide can be obtained using solvents such as water, alcohol and other aprotic solvents combined with sonication or stirring. [58] Using reducing agents, such as hydrazine, [41, 72, 73] dimethylhydrazine, [49] sodiumborohydride followed by hydrazine, [74] hydroquinone, [75] and UV-irradiated $TiO_2$, [76] colloidal graphene oxide can be further chemically reduced producing chemically reduced graphene. Chemical reduction of graphene oxide provides an efficient route for production of graphene. But the hazardous nature and cost of the chemicals used in reduction may limit its application. [58] On the other hand, thermally reduced graphene oxide (RGO) can be produced by rapid heating of dry graphite oxide under inert gas and high temperature. [77-79] For example, heating graphite oxide in an inert environment at 1000 $^{o}$C for 30 s leads to reduction and exfoliation of graphite oxide, producing thermally reduced graphene layers. Compared with chemical reduction methods, the thermal reduction methods can produce chemically modified graphene layers without the need for dispersion in a solvent. [58]

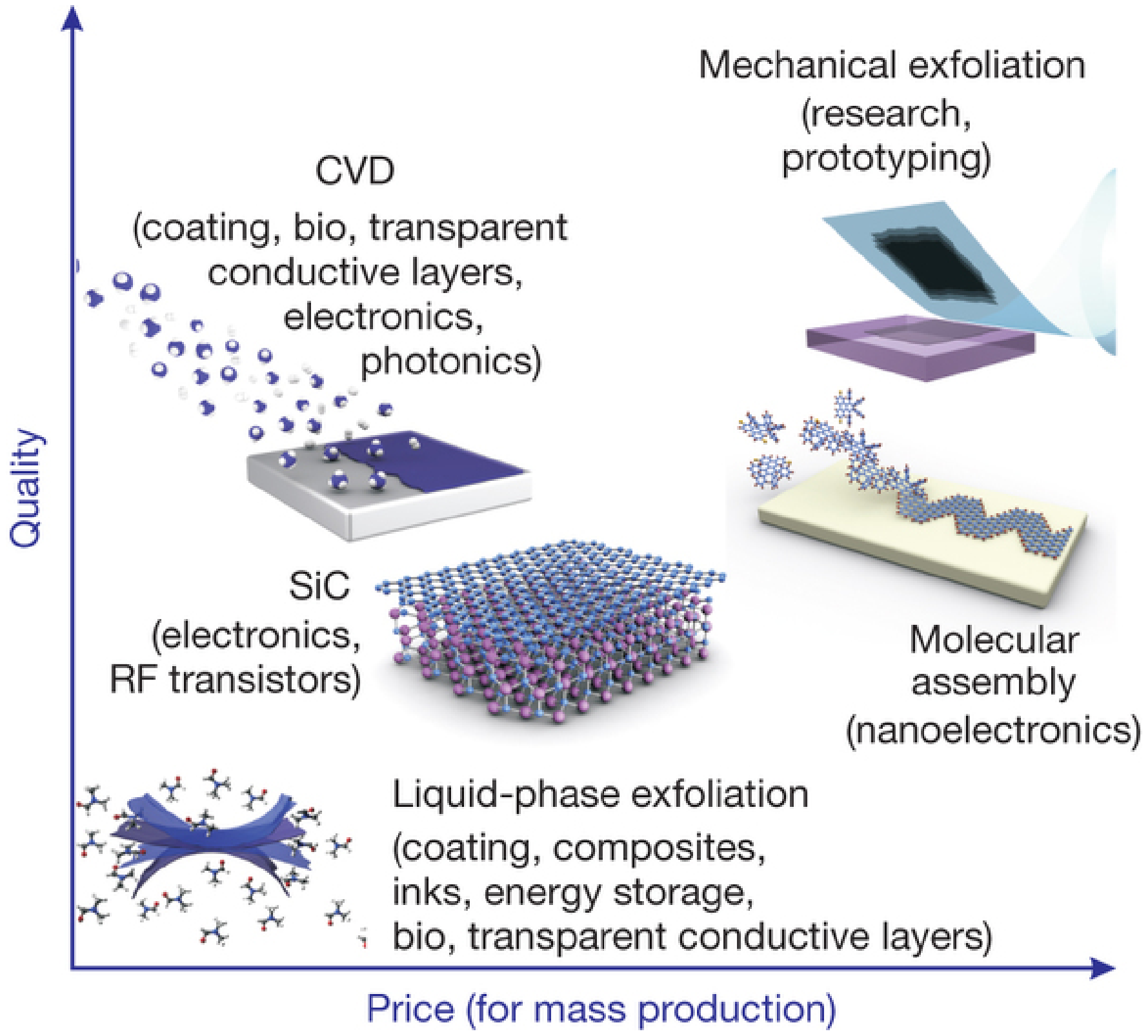

**Figure 3.** Methods of mass production of graphene. [56]

## 2.2 Preparation of graphene/polymer nanocomposites

The dispersion of graphene and its derivatives in polymer is the most important factor for the fabrication of graphene/PNCs. If homogeneous dispersion of graphene can be achieved, the gas barrier performance of graphene/PNCs can be improved significantly. Numerous attempts have been made to improve graphene exfoliation in polymer matrices. [80-94] Most graphene/polymer nanocompoistes have been developed using three strategies: solution mixing, melt mixing and *in situ* polymerization.

Filler that has good compatibility with polymer matrix usually reduces the permeability, mainly because of the reduction of the transport cross section and the increase in the tortuous paths for gas molecules. [95] It is well known that polymer and graphene have poorly compatible. The polymer chains could not tightly contact the graphene nanosheets. Thus, it forms a narrow gap surrounding the graphene nanosheets. Meanwhile, graphene tend to aggregate (layer by layer stacking) due to the poor compatibility of the graphene and the polymer matrix. The gas molecules will be flow through the narrow gap or the relatively highly permeable polymer matrix rather than graphene. Therefore, the PNCs exhibits a low gas barrier performance with decreased diffusional pathways. Surface modification of graphene with organic modifying agents is one suitable choice to solve this problem. [96] The functional groups present on the graphene nanosheet surface can improve the compatibility of graphene with the polymer matrix, which leads to well dispersion and exfoliated polymer/graphene nanocomposites with same volume fraction of graphene. It has been also reported that functionalization to enhance compatibility of the graphene with the polymer matrix may help to moderate the composite solution viscosity, which is benefit for the preparation of PNCs by solution mixing method. [97]

### 2.2.1 Solution mixing

Solution mixing method has been widely used as an effective technique for fabrication of graphene/PNCs due to the ease of processing graphene and its derivatives in water or organic solvents. [98-103] This process generally involves mixing of colloidal suspensions of graphene-based materials with the desired polymer, either itself already in solution or by dissolving the polymer in the same solvent used for filler dissolution, by simple stirring or shear mixing. [104, 105] The main advantage of this method is that it allows the synthesis of PNCs with low or even no polarity. [19] Sonication is often practiced to get better dispersion of graphene nanofillers. Because unmodified graphene has very limited solubility in most organic solvents, [52] functionalization of graphene is needed to improve the solubility. [106-108] Modified graphene can be dispersed easily in a suitable solvent, such as water, acetone, chloroform, tetrahydrofuran, dimethyl formamide or toluene. [19] After mixing is complete, the solvent is needed to remove completely by evaporation or distillation and the polymer material containing the fillers is then molded to give shape to the composite. [38, 109] It has been reported that organic solvents are strongly absorbed on the graphitic galleries of graphite oxide. [110] The crucial challenges in solution mixing are minimizing the residual solvents [38] and obtaining good dispersion properties of the fillers in viscous polymeric solutions. [4] Although the solution mixing approach generally leads to better particle dispersion than melt mixing process, slow solvent evaporation often induces particle reaggregation. In addition, the use of large amounts of solvent and the associated environmental pollution have prevented this technique on a mass scale for the fabrication of graphene/PNCs. [109, 111].

### 2.2.2 Melt mixing

Melt mixing is a typical method for the preparation of thermoplastic PNCs. [112] In this technique, no solvent is required and graphene or its derivatives is mixed with the polymer matrix in the molten state. A thermoplastic polymer is mixed mechanically with graphene or its derivatives at elevated temperatures using conventional methods, such as extrusion and injection molding. [113-115] Traditional mixing equipment such as extruder, internal mixer, and two-roll mill can be used for the melt mixing operations and are usually available in most compounding units. [109] The polymer chains are then intercalated or exfoliated to form nanocomposites. [19] In the melting process, no solvent is involved during the preparation of the composite, [116] making it as an economical and environmentally friendly method for mass production in industrial applications. [9, 58, 116] In this process, strong shear forces are required to blend the highly viscous molten polymers and graphene or its derivatives. The high viscosity of the material often causes non-uniform dispersion of graphite platelets. [17, 98, 99, 117-119] Although the melt mixing process is the most economically viable and environmentally friendly approach, it is not a suitable option for matrices or inclusions that are prone to thermal degradation. [106] Because of thermal instability of most chemically modified graphene, which used for PNCs by melt mixing has so far been limited with thermally stable graphene, such as thermally reduced graphene (TRG). [58] Another challenge for melt mixing is the low bulk density of graphene, which makes extruder feeding a troublesome task. In addition, graphene buckling may cause rolling or shortening of the graphene derivatives. [4]

### 2.2.3 *In situ* polymerization

*In situ* polymerization is another efficient route for the preparation of homogeneously dispersed graphene or its derivatives in a polymer matrix without a prior exfoliation step. [109, 112] In this process, uniformly dispersed graphene nanosheets in solvent can be mixed with a monomer (and/or oligomer) solution with an initiator (e.g., photoinitiators and thermal initiators). After the initiator is dissociated by radiation or thermal energy, [19, 120, 121] exfoliated graphene nanosheets can be mixed or cross-linked with polymer chains and the *in situ* compounding technique confers strong interactions between the filler and the polymer via chemical bonding. [109] It was found that the *in situ* polymerization process is not suitable if the polymerization takes place at the surface of graphene or between graphene layers. [122, 123] During polymerization, however, the viscosity usually increases, which may reduce the processability of nanocomposites. [38, 58] Compared with melt and solvent mixing methods, *in situ* polymerization can provide enhanced dispersion properties and also better compatibility between graphene and the polymer through the introduction of additional functional groups on the graphene (or its derivatives) surfaces. [4] *In situ* polymerization should also be performed in the solution state; therefore, elimination of residual solvents should also be addressed. [4, 112] Moreover, this method requires monomer unites and lot of reagent for the polymerization procedure, and thus less applicable in the case of naturally existing polymers. [17]

In addition to these three common methods, many other methods are also used to prepare graphene/PNCs. For example, layer-by-layer (LbL) assembly process has been proposed as a feasible route to construct alternating layers of nanoplatelets and polymer and, subsequently, to impart the parent polymer with markedly enhanced gas barrier properties. [124-128]

## 3. Barrier properties of graphene/polymer nanocomposites

It has been reported that defect free graphene is impermeable even to Helium. [53, 129] Due to its high aspect ratio and high electronic density of the carbon rings, graphene is able to repel atoms and gas molecules and therefore has a very low solubility to gases. Meanwhile, graphene is also impermeable for biological cell, water and proton diffusion. For instance, Mohanty *et al.* demonstrated that bacterial cells can be encased within a graphenic chamber by wrapping them with protein-functionalized graphene. The strongly repelling π clouds in the interstitial sites of graphene's lattice reduces the graphene-encased-cell's permeability from 7.6-20 nm/s to practically zero. [130] Xu *et al.* used graphene as an atomically flat coating for atomic force microscopy to determine the structure of the water adlayers on mica. The graphene coating can tightly seal and stably "fix" the water adlayer structures on the surface of mica. [131] Algara-Siller *et al.* also report that water can be locked between two graphene sheets to form 'square ice', a phase having symmetry qualitatively different from the conventional tetrahedral geometry of hydrogen bonding between water molecules. [132] Hu *et al.* reported that monolayers of graphene are highly permeable to thermal protons under ambient conditions, whereas no proton transport is detected for bilayer graphene. [133] Achtyl *et al.* placed single-layer graphene on top of a fused silica substrate to cycles of high and low pH. They found that protons transferred reversibly from the aqueous phase through the graphene to the other side where they underwent acid-base chemistry with the silica hydroxyl groups. [134]

**Table 1** Gas permeability of graphene/polymer nanocomposites

| Polymer | Filler | Filler Loading | Processing $^{a}$ | Gas | Permeability $^{b}$ | Reduction |
|---|---|---|---|---|---|---|
| PLA[6] | GONS | 1.37 vol% | Solution | $O_2$<br>$CO_2$ | $1.145\times10^{-14}$ $cm^3$ cm $cm^{-2}$ $s^{-1}$ $Pa^{-1}$<br>$1.293\times10^{-14}$ $cm^3$ cm $cm^{-2}$ $s^{-1}$ $Pa^{-1}$ (50%RH) | 45%<br>68% |
| Cellulose[18] | GNPs | 5 wt% | Solution | $O_2$<br>$CO_2$ | ~0.8 $\times10^{-18}$ $m^3$ $m/m^2$ s Pa<br>~$1.15\times10^{-18}$ $m^3$ $m/m^2$ s Pa | ~27%<br>~34% |
| PS[61] | Graphene | 2.27 vol % | Solution | $O_2$ | 1.84 barrer | 61% |
| IIR[62] | TRG | 5 phr | Solution | $O_2$ | 28.4 $ml/m^2/24h$ | 26% |
| PMMA[67] | Graphene oxide | 1 wt% | Solution | $O_2$ | ~1.25ml/[$m^2$.day.atm] (RH=50%) | 50% |
| HDPE[96] | DA-GO, DA-RGO | 1 wt% | Solution | $O_2$ | $1.75\times10^{-14}$ $cm^3$ cm/($cm^2$ s Pa) | 67% |
| LLDPE[135] | DA-G | 1 wt% | Solution | $O_2$<br>$N_2$ | 19.5 fm/Pa.S<br>5.7 fm/Pa.S | 47%<br>52% |
| PLA[136] | Graphene oxide, GNP | 0.4 wt% | Solution | $O_2$<br>$N_2$ | $1.2\times10^{-18}$ $m^2$ $s^{-1}$ $Pa^{-1}$<br>$0.25\times10^{-18}$ $m^2$ $s^{-1}$ $Pa^{-1}$ | 68%<br>77% |
| PET[137] | fGO | 3 wt% | Solution | $O_2$ | $1.14\times10^{-3}$ barrer | 97.4% |
| PVA[138] | GONS | 0.72 vol% | Solution | $O_2$ | $0.24\times10^{-15}$ $cm^3$ cm $cm^{-2}$ $s^{-1}$ $Pa^{-1}$ | 98.9% |
| EP[139] | $Fe_3O_4$/GNPs | 1 wt% | Solution | He | $2.2\times10^{-6}$ Pa $m^3$/s | 94% |
| PU[140] | Graphene oxide | 1 wt% | Solution | He | 0.970±0.027 barrer | 79% |
| PANI[141] | Graphene | 0.5 wt% | Solution | $O_2$ | 0.1056 barrier | 86% |
| PND[142] | AFG | 5 wt% | Solution | $O_2$ | ~25 cc/$m^3$ day | 79.3% |
| PPC[143] | EFG | 5 wt% | Solution | $O_2$ | 51.8 $cm^3$ $m^{-2}$ $day^{-1}$ | 45% |
| EVOH[144] | TRG | 0.5 wt% | Solution | $O_2$ | $8.517\times10^{-15}$ $cm^3$ cm $cm^{-2}$ $s^{-1}$ $Pa^{-1}$ | 99.98% |
| PVA[145] | Graphene oxide, RGO | 0.3 wt% | Solution | $O_2$ | $5.14\times10^{-15}$ mol $s^{-1}$ $m^{-1}$ $Pa^{-1}$ (60% RH) | 99% |
| TPU[106] | iGO, TRG | 3 wt% | Solution, *In situ*, Melt | $N_2$ | N/A | 98.7% |
| PVA[146] | Graphene oxide | 0.07 vol% | Solution, Recrystallization | $O_2$ | <$5.0\times10^{-20}$ $cm^3$ cm $cm^{-2}$ $Pa^{-1}$ $s^{-1}$ | >99.94% |
| PET[147] | GNPs | 1.5 wt% | Melt | $O_2$ | 0.1cc/$m^2$/day/atm | 99% |
| PU[148] | RGO | 2 wt% | Melt | $O_2$ | $4.13\times10^{-12}$ $cm^3$(STP)/cm-s-cmHg | 90.5% |
| PEN[149] | Graphite & FGS | 10 wt%<br>4 wt% | Melt | $H_2$ | 0.76 barrer<br>0.61 barrer | 47%<br>57% |
| PP[150] | Exfoliated GNPs | 3 vol% | Melt | $O_2$ | ~175 cc mil/m$^2$ atm | ~20% |
| Nylon[151] | FG | 0.3 wt% | Melt | $O_2$ | 10.1 cc/(m$^3$ day) (100% RH) | 47% |
| PC[152] | FGS | 3 wt% | Melt | He<br>$N_2$ | 8.8 barrer<br>0.20 barrer | 30%<br>45% |
| PS[153] | Graphene oxide | 2 wt% | *In situ* | $O_2$<br>$N_2$ | 2.24 barrier<br>0.43 barrier | 25%<br>59% |
| PEI[154] | RGO | 12.5 wt% | *In situ* | $H_2$ | ~46 cc/m$^2$ d atm | 86% |
| PI[155] | RGO | 30wt% | *In situ* | $O_2$ | 26.07 cm$^3$ cm$^{-2}$ 24 h$^{-1}$ atm$^{-1}$ | 93% |
| PMMA[156] | Graphene | 0.5 wt% | *In situ* | $O_2$ | 0.81 barrer | 70% |
| PAN[157] | Expanded graphite | 4 wt% | *In situ* | $O_2$ | ~0.27 Lit/cm$^2$/min | 92% |
| BPEI[36] | Graphene oxide | N/A | LbL | $O_2$ | <0.05 cm$^3$ m$^{-2}$ day$^{-1}$ | 99.6% |
| PEI[158] | Graphene oxide | N/A | LbL | $O_2$ | 0.05 cc/m$^2$ day | 99.4% |
| PDDA, SPVDF[159] | Graphene oxide | N/A | LbL | $H_2$ | 3.1 cc mm/m$^2$ d atm | 92% |
| PEI[160] | Graphen oxide | 91 wt% | LbL | $O_2$<br>$H_2$ | 0.12 cc/m$^2$. atm. Day<br>158.1 cc/m$^2$. atm. Day | 99%<br>41% |
| | | | | $CO_2$ | <1 cc/m$^2$. atm. Day | 97% |
| XNBR[111] | Graphene oxide | 1.9 vol% | Latex co-coagulation | $O_2$ | ~$1.5\times10^{-17}$ m$^2$ Pa$^{-1}$ s$^{-1}$ | 55% |
| SBR[161] | Graphene | 7 phr | Latex compounding | $O_2$ | N/A | 87.8% |

[a] melt, solution and *in situ* in this column refers to melt mixing, solution mixing and *in situ* polymerization, respectively.

[b] 1 barrer= $10^{-10}$ cm$^3$(STP) cm/cm$^2$ s cmHg [137, 162-165]=$7.5005\times10^{-18}$ m$^2$/s Pa [166, 167]

The properties of PNCs depend strongly on how well the fillers are dispersed. The influence of graphene and its derivatives on the barrier properties of PNCs prepared by different processing methods is given in Table 1. It should be noted that the last column in the Table 1, shows the reduction in percentage of permeability of PNCs obtained by different routes. From a theoretical point of view, the presence of impenetrable graphene nanosheets, which are homogeneously dispersed in the polymer matrix, leads to an increase of the diffusion path (tortuosity) and consequently a decrease of the gas permeability of the graphene/polymer composites. [168] The barrier properties of graphene/polymer composites is affected strongly by the aspect ratio, dispersion and orientation of the graphene nanosheets, the graphene nanosheets/polymer interface and the crystallinity of the polymer matrix. [150]

When bulk graphite is exposed to strong oxidizers (such as sulphuric acid, nitric acid, or potassium permanganate), it can be exfoliated into expanded graphite. [169] The scalability and low cost of this process make expanded graphite attractive for industrial applications and the enhanced processability of expanded graphite allows it to be incorporated into polymer matrices. [157] Kalaitzidou *et al.* investigated the effect of exfoliated graphite nanoplatelets (GNPs) on the permeability of PP composites, fabricated by melt mixing using a twin-screw extruder followed by injection molding, as a function of exfoliated GNPs concentration and aspect ratio. The large aspect ratio of exfoliated GNPs, even at low loadings of 3 vol%, increases the oxygen barrier of PP at least 20%. [150] Kim *et al.* prepared polymer nanocomposties reinforced with graphite platelets and functionalized graphite sheets prepared by partial pyrolysis of graphite oxide. Dispersion of unexfoliated graphite and functionalized graphite sheets in poly(ethylene-2,6-naphthalate) (PEN) was explored with melt mixing process. Hydrogen permeability of PEN with 4 wt % functionalized graphite sheets was decreased by 60% while the same amount of graphite reduced permeability only 25%. [149] Expanded graphite was exfoliated on the polyacrylonitrile (PAN) matrix through *in situ* emulsion sonication technique by Prusty *et al*. The permeability of PAN/EG nanocomposites reduced by ~13 times with increasing EG content up to 4 wt%. [157] As mentioned before, the gas barrier properties of nanocomposites depend not only on the loadings of nanoplatelets, but also on their alignments. Jiao *et al.* prepared highly-ordered magnetic graphite nanoplatelets (m-GNPs)/epoxy resin (EP) composites via coating the GNPs with magnetic nanoparticles ($Fe_3O_4$), and then aligning the modified GNPs in EP using low magnetic field (40 mT). $Fe_3O_4$ nanoparticles were tethered onto the surface of GNPs by wet-chemical coprecipitation method. From the scanning electron microscopy (SEM) images of composites, it was confirmed that the m-GNPs in EP matrix have aligned parallel to the direction of the magnetic field (see Figure 4). For the ordered m-GNPs/EP composites prepared with 1.0 wt% m-GNPs, the gas barrier properties improved an order of magnitude compared to EP and more than 65% compared to the randomly dispersed GNPs /EP composites. [139] Al-Jabareen *et al.* prepared PET/GNPs nanocompsoties by melt mixing process. In the best case, the oxygen permeation of the nanocomposites was reduced by more than 99% with 1.5 wt% GNPs. The authors suggested that the GNPs have a two-fold effect on oxygen permeability of nanocomposites, generated simultaneously by their inherent barrier properties and by inducing higher degrees of crystallinity. [147]

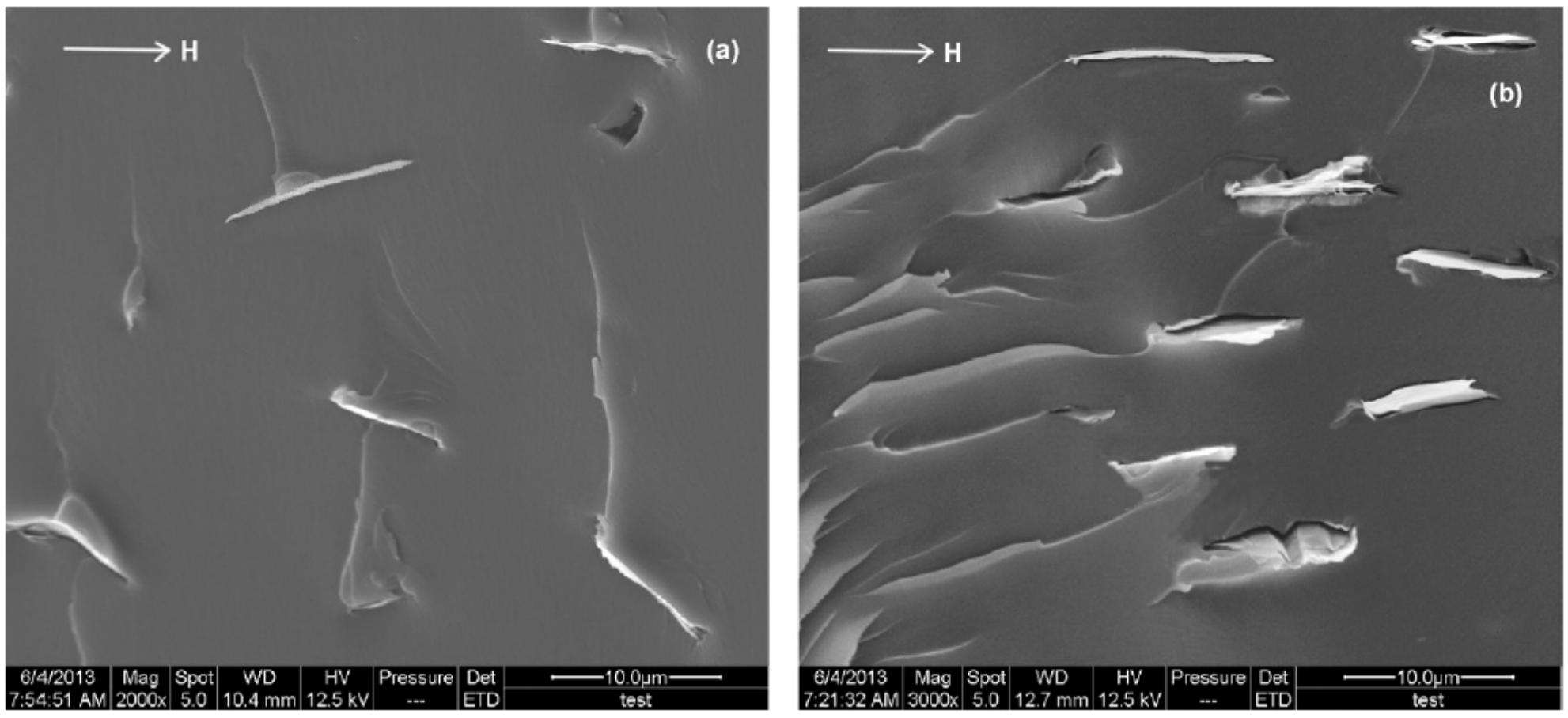

**Figure 4.** SEM images of GNPs/EP and m-GNPs/EP composites. (a) GNPs/EP composite with randomly alignment; (b) m-GNPs/EP composite with high ordered. [139]

On the other hand, graphene oxide can be prepared in bulk quantities by oxidation of graphite with strong oxidants containing many hydrophilic groups, such as hydroxyl, epoxy, and carboxyl acid. [137] Graphene oxide possesses the desirable characteristics of aqueous solution processability attributed to the oxygen-containing functional groups on the basal planes and edges of graphene. [67] These oxygen-containing functional groups promote complete exfoliation and homogeneous dispersion of graphene oxide sheets in polar polymer matrix and improves the interfacial bonding significantly. [6] Graphene oxide has been compounded with various polymers. [170-173] For example, Kang *et al.* fabricated carboxylated acrylonitrile butadiene rubber (XNBR)/graphene oxide nanocomposites which have high mechanical and gas barrier properties using a simple and environment-friendly latex co-coagulation method. The addition of 1.9 vol% of graphene oxide reduced the gas permeability coefficient of XNBR by 55%. [111] Zhu *et al.* prepared graphene oxide from graphite by using a modified Hummers method and then used graphene oxide as a nanofiller to synthesize polyimide (PI)/graphene oxide by *in situ* polymerization (see Figure 5). As expected, the oxygen transmission rate (OTR) decreased significantly from 377.78 of pure PI to 26.07 $cm^3$ $m^{-2}$ 24 $h^{-1}$ $atm^{-1}$ for 30 wt% graphene oxide loaded composite, which displayed a 93% reduction compared with pure PI. [155]

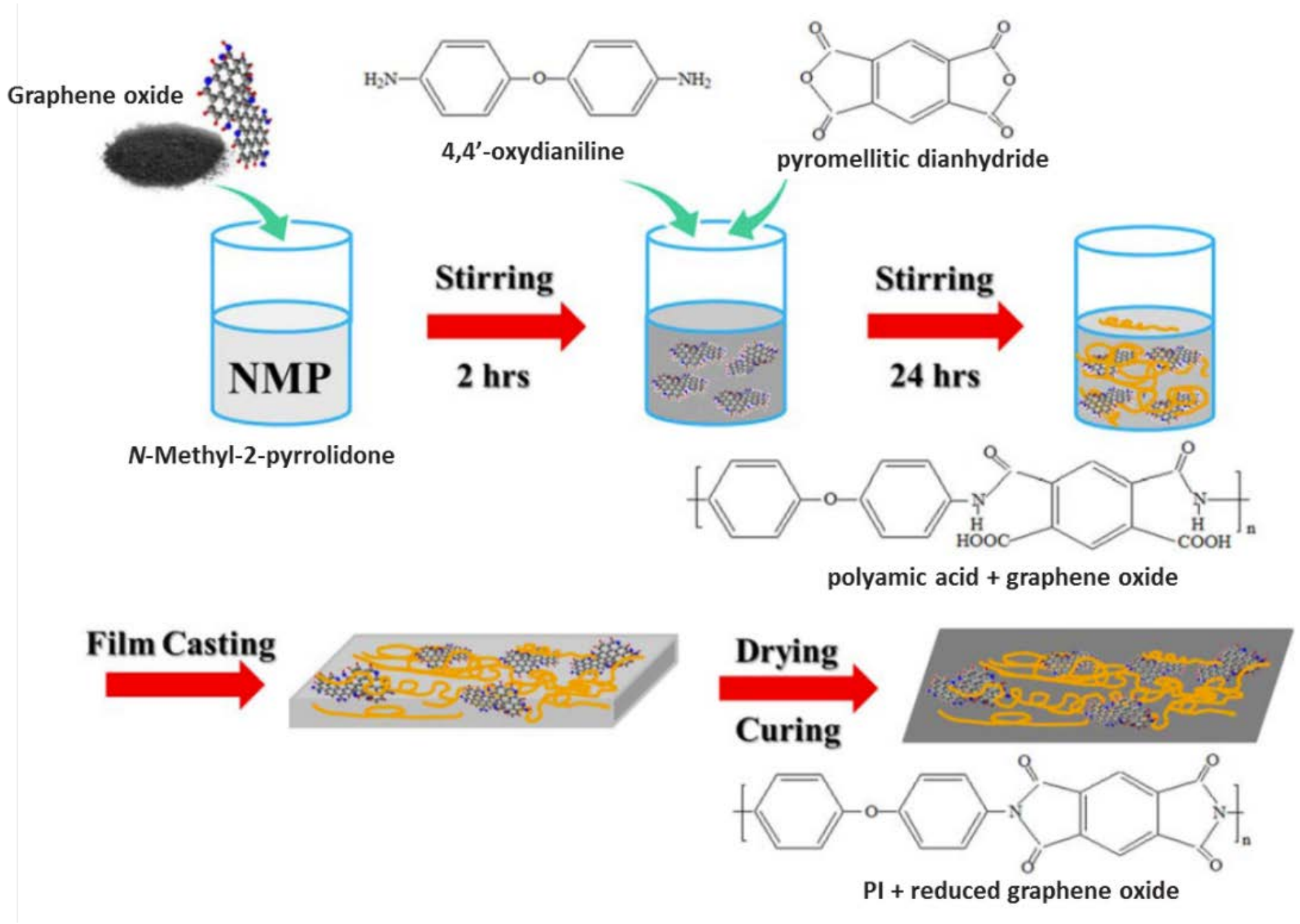

**Figure 5.** The procedures of the synthesis of pure polyimide (PI) and its composite films. [155]

Huang *et al.* prepared high barrier poly(vinyl alcohol) (PVA)/graphene oxide nanosheet nanocomposite films by solution mixing. A more than 98% decrease in the $O_2$ permeability coefficients of PVA film from $21.17 \times 10^{-15}$ to $0.24 \times 10^{-15}$ $cm^3$ cm $cm^{-2}$ $s^{-1}$ $Pa^{-1}$ is achieved by adding only 0.72 vol% graphene oxide nanosheets. The reduction of $O_2$ permeability of PVA film was attributed to excellent impermeable property of graphene oxide nanosheets, their full exfoliation, uniform dispersion and high alignment in the PVA matrix and the strong interfacial adhesion between graphene oxide nanosheets and PVA matrix. [138] Using amidation reaction and chemical reduction, dodecyl amine (DA) functionalized graphene oxide (DA-GO) and dodecyl amine functionalized reduced graphene oxide (DA-RGO) were produced by Ren *et al.* Then, high-density polyethylene (HDPE)/DA-GO and HDPE/DA-RGO nanocomposites were prepared by solution mixing method and hot-pressing process. The crystallinity, dynamic mechanical, gas barrier, and thermal stability properties of HDPE were significantly improved by the addition of DA-GO or DA-RGO. However, the performance of HDPE nanocomposites reinforced with DA-GO was almost the same as that of DA-RGO, which indicated that the reduction of DA-GO was not necessary and the interfacial adhesion and aspect ratio of graphene layers had hardly changed after reduction. [96] Morimune *et al.* developed an environmentally friendly technique for fabricating poly (methyl methacrylate) (PMMA)/graphene oxide nanocomposites in which PMMA was polymerized by soap-free emulsion polymerization and incorporated with graphene oxide using water as a processing medium. The addition of 1% w/w of graphene oxide to the PMMA matrix decreased the permeability by 50% and the nanocomposite with 10% w/w of graphene oxide was found to be almost completely impermeable (see Figure 6). [67]

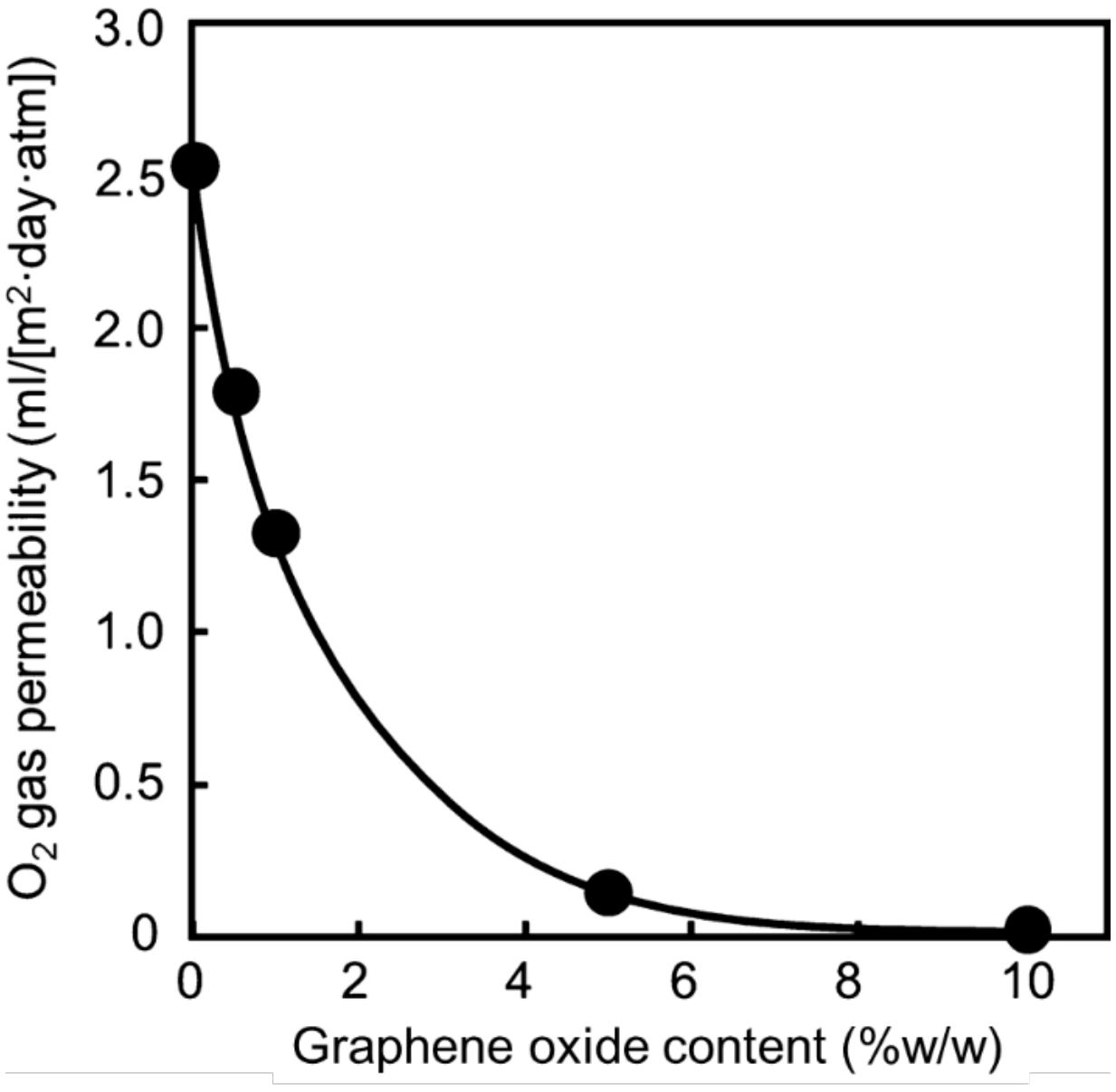

**Figure 6.** $O_2$ gas permeability of PMMA film and PMMA/graphene oxide nanocomposites. [67]

As mentioned before, Graphene oxide prepared by oxidation of graphite with strong oxidants contains many hydrophilic groups. Shim *et al.* reported a method to efficiently convert these hydrophilic groups into alkyl and alkyl ether groups by a one-step reaction of bimolecular nucleophilic substitution with alkyl bromide (see Figure 7). The functionalized graphene oxide (fGO) can be homogeneously dispersed as exfoliated monolayers in various organic solvents while it is precipitated in water. The oxygen permeability coefficients of PET/graphene oxide (1 wt%) and PET/fGO (1 wt%) were found to decrease by 38% and 85% as compared to that of neat PET, respectively. [137]

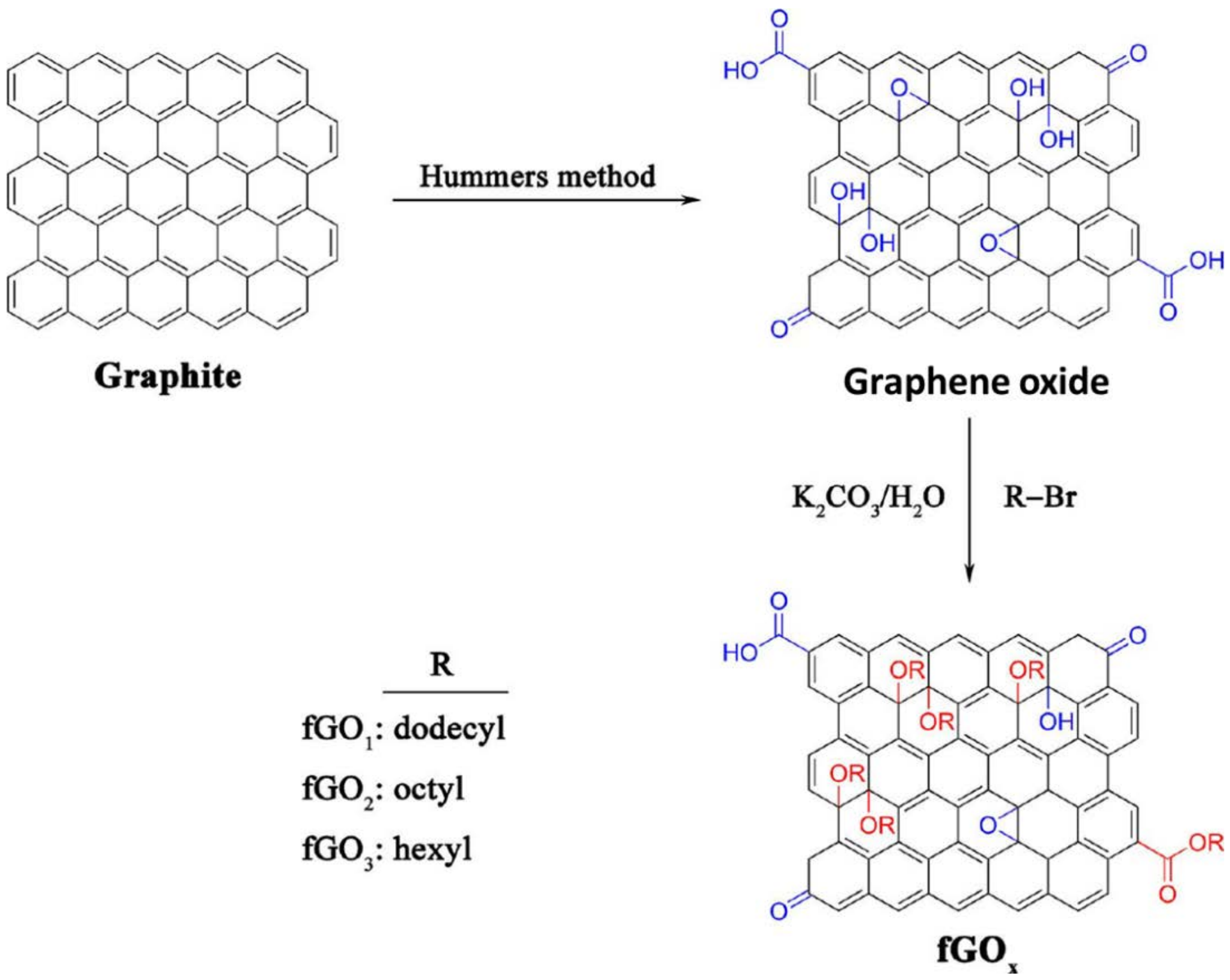

**Figure 7.** Synthetic procedure for functionalization of graphene oxide. [137]

Chen *et al.* synthesized a transparent, gas barrier film comprised of PVA and graphene oxide through combined methods of solution mixing and isothermal recrystallization (see Figure 8). When $O_2$ was passed through the PVA/graphene oxide nanocomposite, it followed a tortuous diffusion path because graphene oxide sheet was impermeable to gases. This reduced the permeability of PVA. However, the $O_2$ molecules could still pass through the spaces between the graphene oxide sheets. Furthermore, isothermal recrystallization was applied to further enhance the barrier property. The graphene oxide acted as a nucleating agent and some PVA crystals were formed around the graphene oxide sheets. In the unique PVA crystal/graphene oxide hybrid structure, the newly formed PVA crystals fill in the spaces between the graphene oxide sheets and the graphene oxide sheets were linked together by PVA crystals as a bridge. When $O_2$ was passed through PVA crystal/graphene oxide hybrid structure, it was forced to follow a longer diffusional pathway, leading to an ultra-low permeability of the composite. The recrystallized PVA/graphene oxide film with only 0.07 vol% graphene oxide showed an $O_2$ permeability $<5.0\times10^{-20}$ $cm^3$ cm $cm^{-2}$ $Pa^{-1}$ $s^{-1}$. The high $O_2$ barrier properties were attributed to the unique hybrid of PVA crystals and graphene oxide sheets. [146]

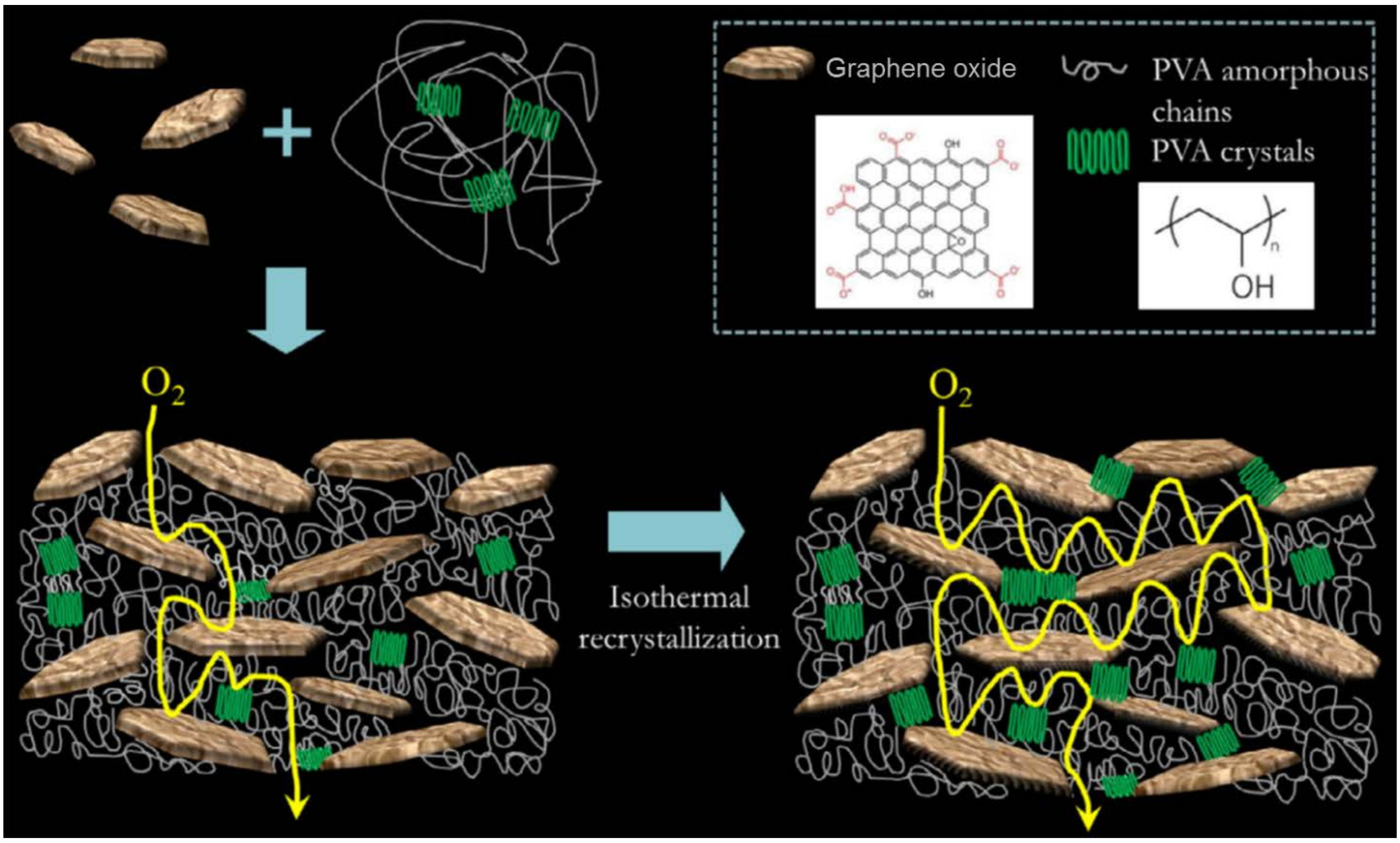

**Figure 8.** Formation of gas barrier film of PVA/graphene oxide with hybrid structure. [146]

Liu *et al.* presented a facile approach for the simultaneous *in situ* reduction of graphene oxide, its surface modification and preparation of polyethylenimine (PEI)/RGO nanocomposites with improved gas barrier properties (see Figure 9). PEI was employed both as reducing agent and surface modifier for the preparation of water dispersible graphene as well as a polymer matrix to obtain PEI/RGO composite for the gas barrier application. The gas barrier properties of PEI/RGO increased with increasing PEI content in the composite film. For example, the value of the hydrogen permeation rate for PEI/RGO/PET was around 7 times lower than that of the bare PET substrate when the PEI : graphene oxide feeding ratio was 7:1,. [154] Huang *et al.* fabricated a set of poly(lactic acid) (PLA)/graphene oxide nanosheets (GONS) nanocomposites. GONS were fully exfoliated and randomly dispersed in PLA matrix. The $O_2$ and $CO_2$ permeability coefficients of PLA films were decreased by about 45% and 68% respectively at a low GONS loading of 1.37 vol%. The enhanced gas barrier performance was ascribed to the impermeable property of GONSs as well as the strong interfacial adhesion between GONSs and PLA matrix. [6]

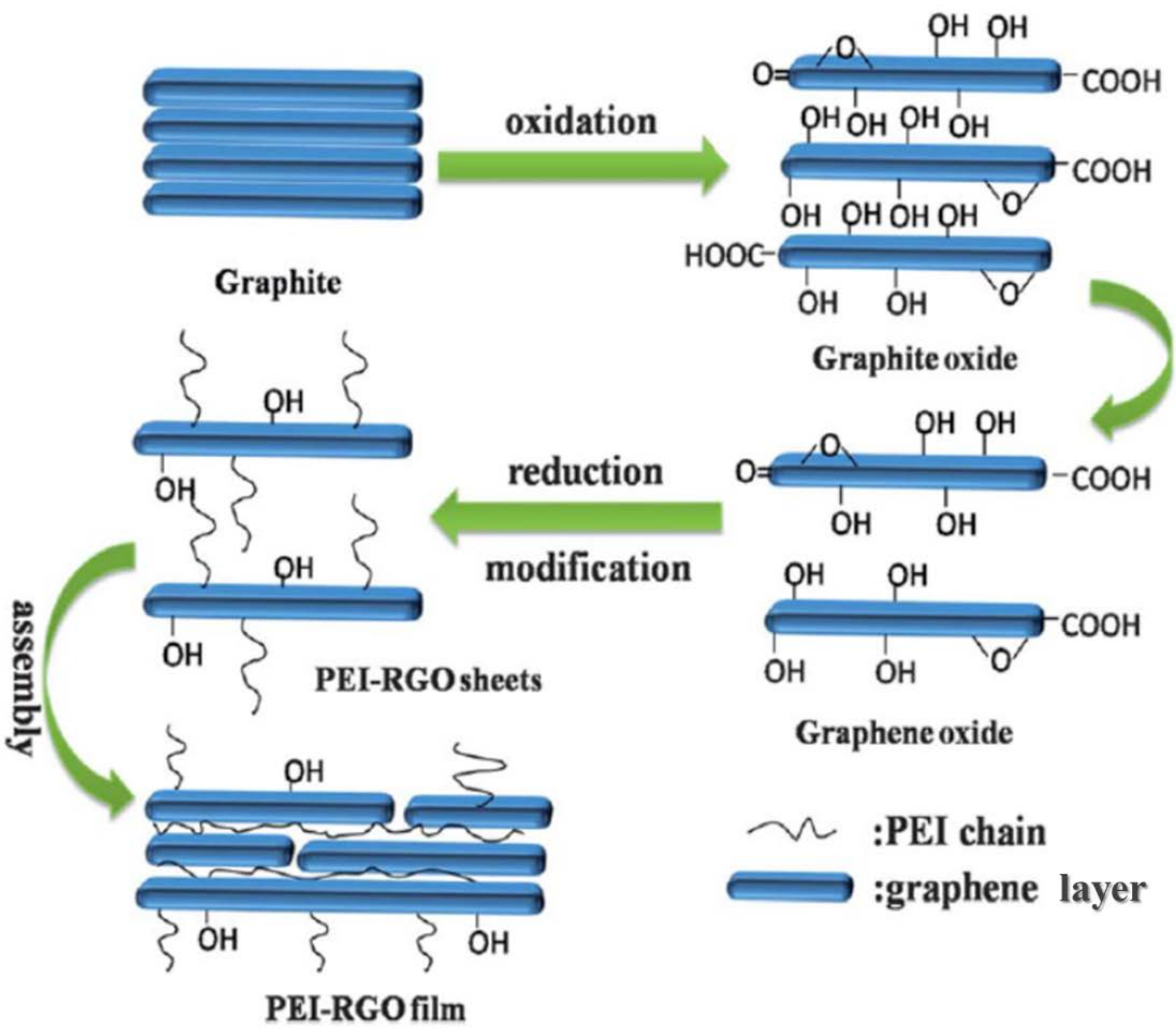

**Figure 9.** Schematic of polyethylenimine (PEI)/RGO synthesis and the assembly process. [154]

Kim *et al.* used different graphene layers exfoliated from graphite oxide via two processes: chemical modification and thermal exfoliation to prepare thermoplastic polyurethane (TPU)/graphene nanocomposites by three different methods of dispersion: solution and melt mixing, *in situ* polymerization (see Figure 10). Transmission electron microscopy (TEM) and wide-angle X-ray diffraction (WAXD) results indicated that solution mixing techniques more effectively distribute thin exfoliated graphene layers in the polymer matrix than melt processing method. They found that *in situ* polymerized thermal reduced graphene oxide is not as effective as solution mixed thermal reduced graphene oxide in reducing gas permeability. Up to 80 fold decrease in nitrogen permeation of TPU was observed with only 3 wt% phenyl isocyanate treated graphite oxide. [106] Jin *et al.* fabricated a series of nylons 11 and 12/functionalized graphene (FG) nanocomposties by melt mixing with pre-mixing. The nylon 11 films with FG loading as low as 0.3 wt% showed a superior reduction of oxygen permeability by ~47%. Aside from the dispersion of graphene and interface between the polymer and graphene, it can be concluded that flexibility of graphene in the polymer matrix is also an important factor to achieve the maximally improved gas barrier properties due to the graphene wrinkled structure in the polymer matrix. [151] The presence of low winkled structure of graphene could provide a larger area of the effective tortuous paths, which could benefit to enhance barrier properties.

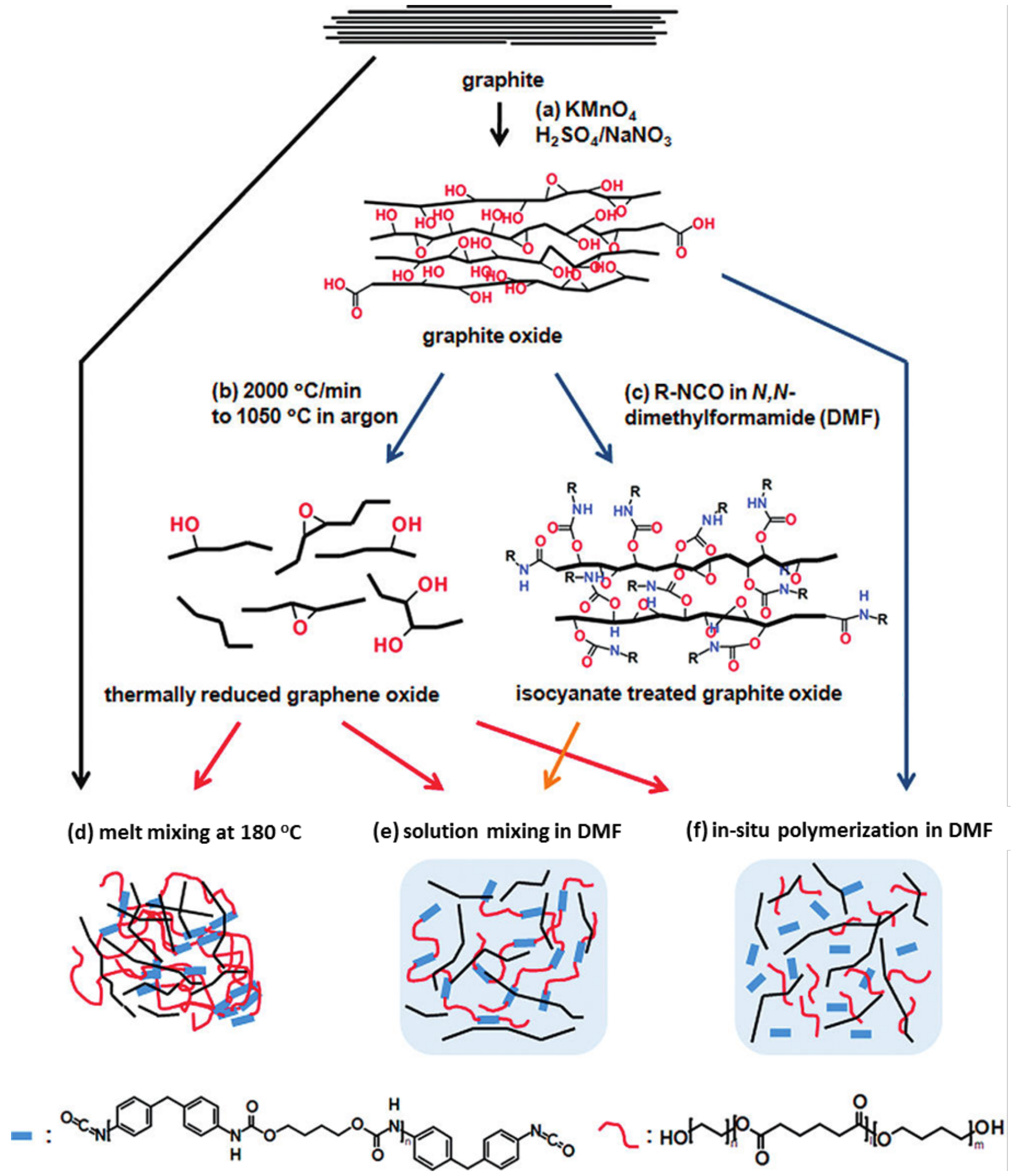

**Figure 10.** Schematics of thermoplastic polyurethane (TPU)/graphene composite preparation routes: (a) after oxidizing graphite in concentrated acids, functionalized graphene layers can be obtained via (b) rapid thermal expansion (thermally reduced graphite oxide) or (c) organic modification with isocyanate (R-NCO) in *N*, *N*-dimethylformamide (DMF) (isocyanate treated graphite oxide); TPU/graphene composite prepared via (d) melt mixing, (e) solution mixing and (f) *in situ* polymerization. (Black lines represent graphitic reinforcements. TPU hard and soft segment are denoted by short blue blocks and thin red curves, respectively) [106]

Compton *et al.* prepared a PS/graphene film with low $O_2$ permeability and reduced transparency. At only 0.02 vol%, crumpled graphene nanosheets significantly densified PS films and thus lowered the free volume within the polymer matrix, which resulted in an unprecedented reduction in oxygen permeability. At such a low concentration of crumpled graphene nanosheets, the PS/graphene films were superior in reducing the $O_2$ permeability of PS to those of some of the best reported polymer/clay nanocomposites with ~40 times more nanofiller loading. [61, 174] Using solution mixing method, Mahmoudian *et al.* presented an approach for the preparation of regenerated cellulose (RC)/GNPs nanocomposites via an ionic liquid, 1-ethyl-3-methylimidazolium acetate (EMIMAc). The nanocomposites exhibited improved gas barrier properties compared to RC and the permeability of $CO_2$ through the nanocomposites was higher than that of $O_2$. It was suggested that the variation of permeability between different gases was depended on the size of the gas molecules and its solubility in polymer matrix. [18, 175]

Recently, LbL assembly process has been applied as a simple and versatile thin-film fabrication technique. [127, 128, 159, 176-188] The thickness, transparency and gas permeability of LbL film can be precisely controlled by tailoring aqueous deposition mixture concentration, [189] temperature, [190] pH [191] and exposure time. [160, 192] Single layers of graphene oxide were alternately deposited with branched polyethylenimine (BPEI) to investigate the oxygen barrier of these thin film assemblies by Yang *et al.* A 10 bilayer film deposited on PET, made with 0.1 wt% BPEI and 0.2 wt% graphene oxide mixtures, reduced the OTR by a factor of 71 relative to an uncoated PET substrate and the OTR of the BPEI/graphene oxide bilayer films was only 0.12 cc $m^{-2}$ $day^{-1}$, which was comparable to a 100-nm $SiO_x$ nanocoating [193] and two orders of magnitude better than a 25-μm ethylene-vinyl alcohol (EVOH) copolymer film. [160, 194] Chen *et al.* also prepared transparent multilayered gas barrier films consisting of BPEI/ graphene oxide on a PET substrate by LbL assembly technique (see Figure 11). They investigated the effect of the graphene oxide suspension pH on the oxygen barrier properties of the BPEI/graphene oxide film. It was demonstrated that the oxygen barrier properties of the multilayer film were depended on the pH of graphene oxide suspension. The BPEI/graphene oxide film prepared using a graphene oxide suspension with a pH of 3.5 exhibited very dense and ordered structures and delivered a minimum OTR value (<0.05 $cm^3$ $m^{-2}$ $day^{-1}$). When the pH value of graphene oxide was lower or higher than 3.5, the BPEI/graphene oxide film showed higher OTR. [36] Rajasekar *et al.* fabricated multi-layered films containing poly(diallyldimethylammonium) chloride (PDDA) and sulfonated polyvinylidene fluoride (SPVDF)-graphene oxide composites through LBL assembly process to enhance the hydrogen gas barrier properties. The hydrogen gas transmission rate of a 16 bi-layer LBL assembly with 2 wt% graphene oxide was 11.7 cc/$m^2$ d atm, which was much lower than that of PET substrate (329.1 cc/$m^2$ d atm) and a one bi-layer LBL assembly without graphene oxide (277.9 $cc/m^2$ d atm).[159] Yu *et al.* prepared a transparent and electrically conductive oxygen barrier film composed of graphene oxide and PEI. Graphene oxide was produced by oxidizing graphite via chemical methods. Alternating graphene oxide layers and PEI layers were deposited on a PET film surface by LbL process. As the number of deposition layers increased, the OTR decreased from 8.229 to less than 0.05 $cc/m^2$ day, the minimum sensitivity of the measurement.[158]

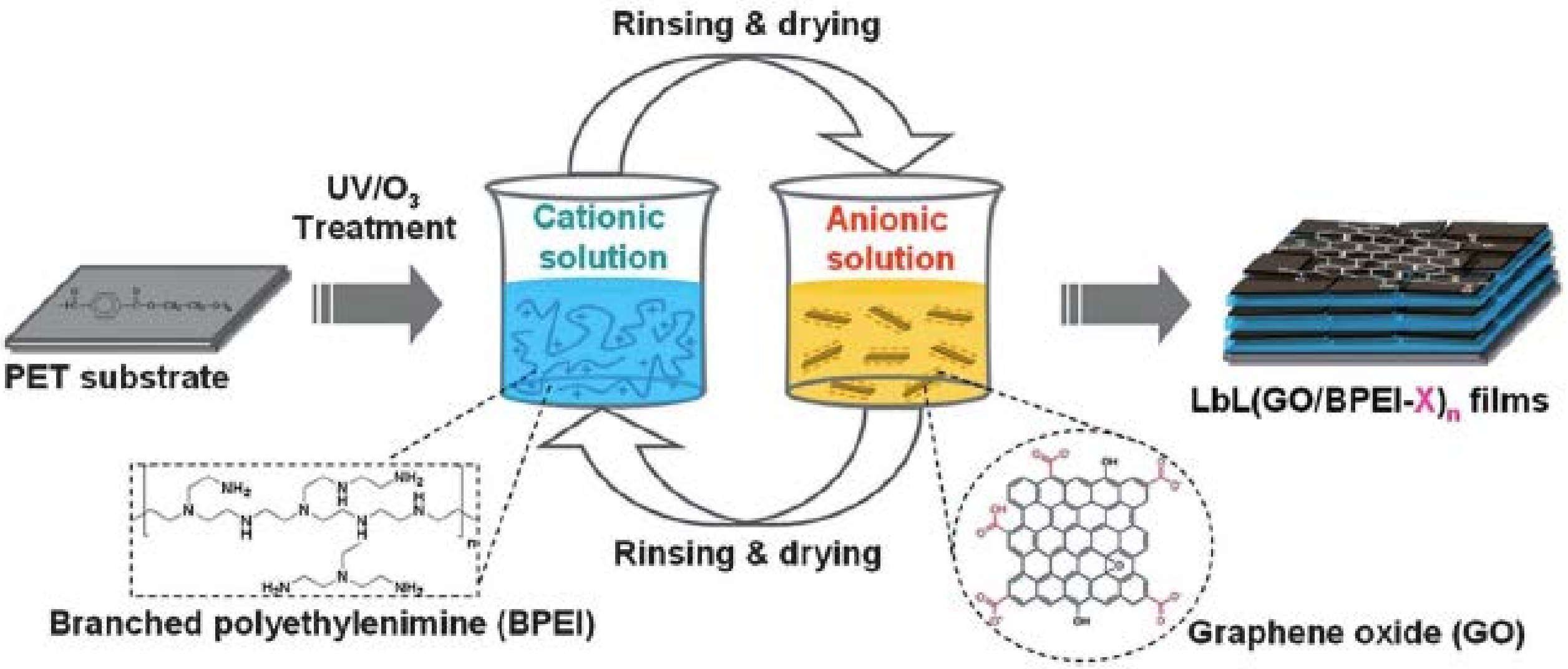

**Figure 11.** Schematic of preparing BPEI/graphene oxide films through LbL assembly.[36]

### 4. Modeling of the barrier properties of graphene nanocomposites

The review on barrier properties of graphene nanocomposites is hitherto presented based on experimental findings. Nonetheless, there are analytical and numerical models that extract barrier properties of graphene nanocomposites. Therefore, in this section, models to determine the barrier properties of graphene nanocomposites are reviewed. The barrier properties of a nanocomposite can be described by three common coefficients: permeability coefficient (P), diffusion coefficient (D) and solubility coefficient (S). The gas permeation in a nanocomposite is governed by a diffusion-solubility mechanism[195] and it occurs due to a pressure gradient across the nanocomposite film. For steady-state diffusion across the film, gas permeability measurements can be performed using the constant-volume, variable-pressure approach.[196] In this approach, vacuum is applied on both sides of a film of effective thickness, $t_m$, situated in the permeation cell and the permeability coefficient is determined. The permeability of gas as a single molecule through a nanocomposite film is generally supposed to be a function of two processes: diffusion and solubility. The diffusion coefficient describes the kinetic aspect of the transport, and the solubility coefficient relates the penetrant affinity and the thermodynamic aspect of the transport. On the basis of Fick's and Henry's law, the relation of gas permeability can be expressed as follows:

$$Q = DSAt\left(\frac{\Delta p}{t_m}\right) \quad (1)$$

where D and S represent the diffusion and solubility coefficients of a gas in the graphene nanocomposite film, respectively; A is the effective area of the film ($nm^2$), Q is the gas leakage quantity through the film ($nm^3$), t is the time of gas permeation (s), and $\Delta p$ is the pressure difference across the film (nm Hg). According to the diffusion-solution model, the permeability in graphene nanocomposite film can be expressed as a product of the diffusivity and solubility as follows:

$$P = DS \quad (2)$$

Equation (2) holds true when the value of D is independent of the concentration and the value of S follows Henry's law, and it is often considered to describe the gas transport properties of composites reinforced with impermeable nanofillers in a polymer matrix. [195, 197] In the above diffusion-solubility model, penetrant molecules initially dissolve into the high pressure side of a film then diffuse across its thickness and finally desorb at the low pressure side. Thus, the permeability of a penetrant depends on both of its diffusivity and solubility, and these properties can be systematically altered through judicious choice of molecular design and environmental factors. Nanocomposite films possess better barrier properties in comparison to homogenous films [198] and the simplest model by Picard *et al.*,[199] given below, can predict the penetrant solubility in the nanocomposite film:

$$S = S_o(1 - \phi) \quad (3)$$

where $S_o$ and $\phi$ denote the penetrant solubility coefficient of the pure polymer matrix and the volume fraction of graphene based nanofillers in the polymer matrix, respectively.

In the nanocomposite film, the dispersed graphene nanofillers act as impenetrable barriers and therefore penetrant follow tortuous pathway in order to diffuse through the film thickness; this increases the effective pathway for diffusion of the gas, thus degrading the diffusion coefficient. This effective diffusion coefficient (D) of nanocomposites can be expressed in the following form by introducing tortuosity factor ($\tau$):

$$D = \frac{D_o}{\tau} \quad (4)$$

in which $D_o$ is the diffusion coefficient of the pure polymer matrix [9] and tortuosity factor is defined as:

$$\tau = \frac{t_m}{t'_m} \quad (5)$$

where $t'_m$ is the distance between tortuous pathways through the nanocomposite film thickness, that is, the shortest pathways for gas molecules. [4, 8] Figure 12 shows a simple gas permeability model for a rectangular array of graphene layers for the case where gas molecules pass through the film in the perpendicular direction.

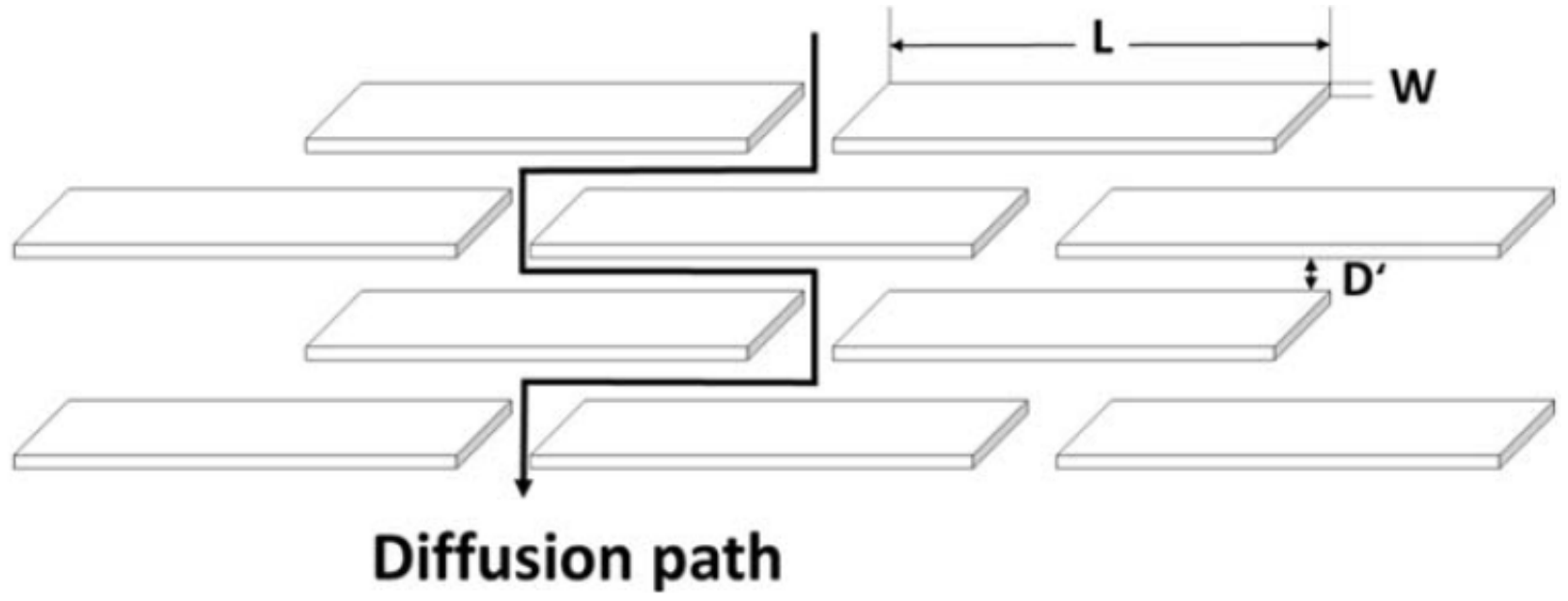

**Figure 12.** Regular arrangement of horizontally staked graphene layers in a parallel array perpendicular to the diffusion direction. [4]

These layered graphene/PNCs structures maximize the gas-diffusion path length and, as a result, significantly decrease the gas diffusion through layered film. The diffusion thickness ($t'_m$), for the average number of N graphene layers dispersed in the film, can be determined as follows: [4]

$$t'_m = t_m + \langle N \rangle \frac{L}{2}; \quad \langle N \rangle = \frac{t_m}{D' + W} \quad (6)$$

where L and W denote the width and thickness of the graphene layer, respectively, and $D'$ is the effective distance between two adjacent graphene layers. Effective permeability coefficient (P) is obtained using Eqs. (2)-(4); as follows

$$\frac{P}{P_o} = \frac{1 - \phi}{\tau} \quad (7)$$

where $P_o$ is the gas permeability of pure polymer matrix. The value of $\tau$ depends on the shape and volume fraction of nanofillers, and several researchers calculated its value for different types of fillers. These studies, however, are concerned only with a rather limited range of composites; but we can make use of these studies to determine the different values of tortuosity factor for graphene/PNCs systems since it is a function of shape and size of the dispersed particles.

Several earlier studies focused on the use of composite theory [see Eq. (3)] to determine the effective permeability in nanocomposites. Some of the pioneering research by Barrer *et al*. on permeation in composites included measurements on rubbery polymers containing inorganic fillers like zinc oxide and silica. [200, 201] Their experimental results were found to be in good agreement with the predictions of composite theory. [201] Some other researchers developed various simple models to determine the gas transport behavior of PNCs filled with inorganic nanoplatelets. [11, 168, 202-205] For instance, Takahashi *et al.* [204] investigated the gas permeation properties of nanocomposites based on butyl rubber with high loadings of vermiculite. In their study, the permeability of the nanocomposite coatings to various gases was measured and validated to ensure the accuracy of permeation models for nanocomposites with flake-like fillers proposed by Cussler, Nielsen, Fredrickson and Bicerano, and Gusev and Lusti (these models are discussed in next section 4.1). Chlorobutyl rubber nanocomposites were prepared by Saritha *et al.* [205] using organically modified cloisite 15 A and characterized using XRD and TEM. The gas barrier properties of the nanocomposites were modeled using the composite theories of permeation and the tortuosity factors were predicted. In their study, the reciprocal tortuosity factors predicted by Gusev and Lusti, and Nielsen permeation models for given values of $\alpha\phi$ of nanoparticles were found to be in good agreement with those of the experimental results; on the other hand, Cussler models show satisfactory agreement with the experimental tortuosity factor at lower values of $\alpha\phi$ but vary differently at higher values; where $\alpha$ is the aspect ratio of nanoparticle. Recently, Yoo *et al.* [4] presented the state-of-the-art research on the use of graphene, GO, and RGO for barrier applications, including few-layered graphene or its derivatives in coated polymeric films. Their analytical results were closely matched with those of experiemntal data.

### 4.1. Analytical modeling of barrier properties

Several empirical models have been proposed to predict the barrier properties of composite materials. These studies considered different contributions to transport, usually related to the "tortuous path" resistance in the nanocomposite system, described in a simplified way. Expressions for widely used empirical models are reported herein without attempting an analysis of assumptions and derivations; the reader is referred to the original papers for more details. Different models, in particular microscale-based analytical models, are presented herein to determine the barrier properties of graphene/polymer composites. We considered different aspects, such as graphene geometry and orientation, graphene-matrix interphase, and agglomeration of graphene layers to model these composites systems. Diffusion of a small solute through a nanocomposite film containing a suspension of impermeable particles is a classic problem in transport phenomena. One of the first examples was Maxwell's theory [206] which predicts the effective permeability for a nanocomposite film containing periodic array of spherical impermeable fillers; as follows

$$\tau = 1 + \frac{1 + \frac{\phi}{2}}{1 - \phi} \tag{8}$$

The Maxwell model assumes that the contact between filler and the surrounding matrix is perfect. It may also be noted that the Maxwell model holds good for a dilute suspension of spherical fillers and may not provide good estimates for the nanocomposite containing graphene layers.

Nanocomposite films which contain impermeable flakes or laminae show permeabilities much lower than conventional membranes, and hence can serve as barriers for oxygen, water and other solutes. [207] At this juncture it may be noted that the Maxwell model is not used herein to predict the barrier performance of nanocomposite containing graphene layers; for the sake of completeness and ready for reference, the authors added the discussion on the Maxwell.

The situation is completely different for a nanocomposite film containing graphene layers oriented perpendicular to the diffusion path. For such case, the value of tortuosity factor is obtained from the following relation [168]

$$\tau = 1 + \alpha\phi \tag{9}$$

The Nielsen approximation (Eq. 9) describes the increase in the tortuosity of the gas diffusion path by relating it to the volume fraction and aspect ratio of the graphene layer. Nielson assumed that the nanoplatelets are completely exfoliated and dispersed along the perpendicular direction of diffusion. It may be noted that Eq. (9) is different from Maxwell's relation because the shape of the particle appears in the tortuosity factor; this indicates that the aspect ratio of particles has significant influence on the diffusion direction. It may also be noted that Eq. (9) can predict the results accurately when the volume fraction (ϕ) of graphene layers is less than 0.01. If the value of ϕ exceeds 0.01 then the graphene layers tend to aggregate in the polymer matrix and Eq. (9) ceases to be valid. Therefore, the maximum value of ϕ is considered as 0.01 to study the effect of aspect ratio of graphene layers. Figure 13 demonstrate the effect of different aspect ratio (α) of graphene layers on the relative permeability ($P/P_o$) of graphene/PNCs. It may be observed from this figure that the barrier performance of nanocomposite significantly improves with the increase in value of α. Significant improvement is observed in the barrier performance of nanocomposite even at low volume fraction of graphene layers of the order of 0.0015 and higher values of α ($\geq 1000$). For the lower values of α ($\geq 50$), Nielsen model predicts a linear barrier improvement with ϕ. It may also be observed that the percentage of reduction in the relative permeability is found to be less for the higher values α ($\alpha \geq 500$) in comparison to lower ones. These results indicate that the graphene and its derivatives can be utilized effectively to improve the barrier performance of nanocomposites.

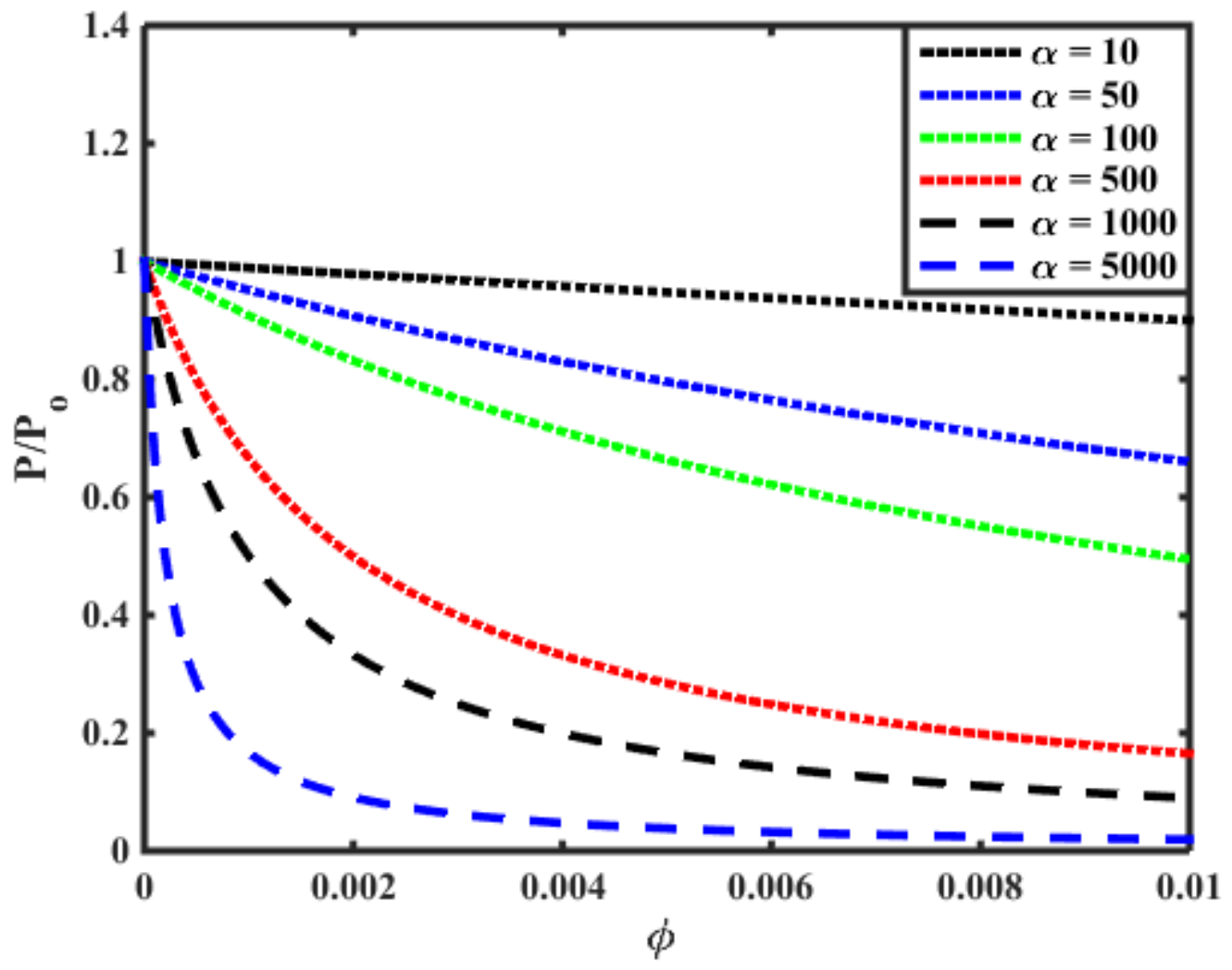

**Figure 13.** Effect of aspect ratio of graphene layers on the barrier performance of nanocomposite.

The experimental data are also found to be in good agreement with those predicted by Eq. (9). For example, as shown in Figure 14(a), the experimentally measured permeability values of graphene/polystyrene films by Compton *et al.* [61] are found to be in agreement with those obtained by Neilsen. Kim *et al.* [145] synthesized PVA/RGO composites using modified Hummers method and a solution-mixing method. In their study, the diffusivity of the PVA/RGO-coated film followed the qualitative feature of the Nielsen approximation.

Nielsen's model is accurate in the dilute regime but it is considered inadequate by Cussler *et al.* [207] when the volume fraction of fillers is semi-dilute (that is, the volume fraction of fillers is low but they overlap each other, $\phi \ll 1$ but $\alpha\phi \gg 1$). In such case, we can use another relation given by Cussler *et al.* for a nanocomposite film containing graphene layers (see Fig. 12) oriented perpendicular to the diffusion: [207]

$$\tau = 1 + \frac{\alpha^2\phi^2}{1-\phi} \tag{10}$$

It is not always possible to fabricate nanocomposite with graphene layers distributed at regular intervals and these sheets may appear randomly in the film. Therefore, in case of (i) two courses of graphene layers with alignment and misalignment occurring with equal probability; and (ii) random misalignment of successive layers of hexagonal graphene layer, Eq. (10) can be written in the following forms, [207] respectively:

$$\tau = 1 + \frac{\alpha^2\phi^2}{2(1-\phi)} \tag{11}$$

$$\tau = 1 + \frac{2\alpha^2\phi^2}{27(1-\phi)} \tag{12}$$

The result for platelets given in Eq. (10) was experimentally verified, especially for barrier films used in packaging. [138, 207-211] Huang *et al.* [138] prepered high barier GONS/PVA nancomposite films and reported that both $O_2$ and water vapor permeability coefficients of PVA film decresed by 98% and 68%, respectively, at a low GONS loading of 0.72 vol%. As deomnstrated in Figure 14(b), it can be observed that their experimental results are close to that of the predicted Cussler values. This finding indicates that GONSs have the similar morphology to that described in the Cussler model, that is, GONSs opt to align parallel to the film surface.

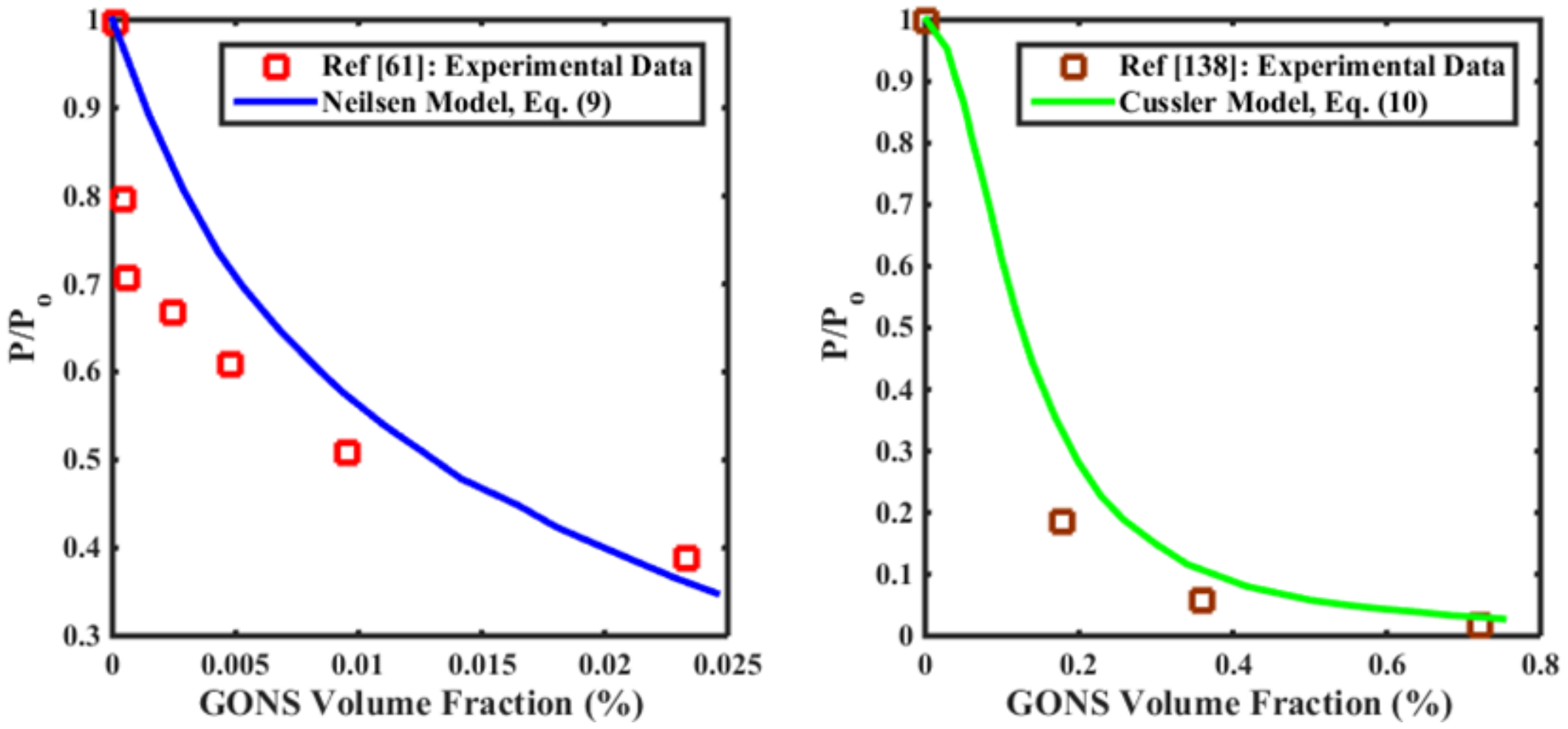

**Figure 14.** Comparison between the experimental data and the results predicted by (a) Neilsen Model [Eq. (9)] and (b) Cussler Model [Eq. (10)] of the nanocomposite films containing GONSs.

Idealizing graphene layers as rectangular platelets of uniform size dispersed at regular intervals in the nanocomposite film, as shown in Figure 12, Eq. (10) can be modified as follows [207]

$$\tau = 1 + \sigma\alpha\phi + \frac{\alpha^2\phi^2}{1-\phi} \tag{13}$$

in which $\phi$ takes different form depending on the range of geometrical parameters of the graphene layer; as follows

$$\phi = \frac{LW}{(L+s)(D'+W)} \tag{14}$$

where $\sigma$ ($= s/W$) is the pore aspect ratio, characterizes the pore shape; and s is the spacing between the adjacent nanoplatelets in the perpendicular direction to diffusion.

Interestingly, two cases of Eq. (13) can be verified experimentally. First, we consider the case where $\alpha\sigma \ll 1$. In this case, the wiggles (this effect is discussed later) within the nanocomposite films are dominant, and Eq. (13) becomes

$$\tau = 1 + \frac{\alpha^2\phi^2}{1-\phi} \tag{15}$$

Second, we consider the case where $\alpha\sigma \gg 1$. Now, Eq. (13) becomes

$$\tau = 1 + \alpha\phi\sigma \tag{16}$$

The above Eq. (16) indicates that the diffusion is limited not by the wiggles but by the nanoplatelets themselves. Equations (15) and (16) provide the desired results for the two-dimensional model shown in Figure 12; they provide the change in the flux caused by the nanoplatelets in the nanocomposite film and this change is a function of three variables, $\alpha$, $\phi$ and $\sigma$, which can be altered experimentally.

Using a conformal mapping method, Aris developed an analytical tortuosity expression for the composite containing two-dimensional periodic array of nanoplatelets. [212] Accounting for the results reported by Aris, Falla *et al.*, [211] proposed the following refined Cussler model:

$$\tau = 1 + \frac{\alpha^2\phi^2}{1-\phi} + \frac{\alpha\phi}{\sigma} + \frac{4\alpha\phi}{\pi(1-\phi)}\ln\left[\frac{\pi\alpha^2\phi}{\sigma(1-\phi)}\right] \tag{17}$$

The first two terms on the right hand side of Eq. (17) are the same as those in Eq. (10), but the third and fourth terms are additional. The physical origin of each of the terms on the right hand side of Eq. (17) merits discussion: (i) the first term is unity and it indicates the pure polymer matrix without the loading of nanoplatelets; (ii) the second term involves $\alpha^2$, is the resistance to diffusion of the torturous path around the nanoplatelets. Such effect is called as “wiggling”. The square $\alpha^2$ and $\phi^2$ terms reflect both the increased distance for diffusion and the reduced cross-sectional area between the nanoplatelets. Wiggling is the main contribution to the increased resistance in platelets-filled barrier films. [207-211] The preferred path for diffusion must be predominantly around the second largest dimension, the short side, of these oriented nanoplatelets; (iii) the third term indicates the resistance of slits between the nanoplatelets; and (iv) the fourth term represents the constriction of a diffusing medium to pass into and out of the narrow slits and this effect is called as "necking". Such effect can be imagined for graphene pierced only by widely separated slits. Solute diffusing across such a layer will be forced to neck down to pass through the slits. This necking down represents an additional resistance to diffusion, even when the slit length is extremely short.

Monte Carlo calculations were carried out by Falla *et al.* [211] to study the diffusion across films containing impermeable flakes. In their study, the effects of tortuous paths around the flakes, of diffusion through slits between flakes, and of constricted transport from entering these slits were investigated. Their calculations show that a simple analytical Eq. (17) developed by Aris [212] reliably predicts these three effects. Both calculations and Eq. (17) show the separate conditions when each of these effects is important.

For the same geometry, as shown in Figure 12, a slightly different model proposed by Wakeham and Mason [213] is given by:

$$\tau = 1 + \frac{\alpha^2 \phi^2}{1-\phi} + \frac{\alpha\phi}{\sigma} + 2(1-\phi)\ln\left[\frac{1-\phi}{2\sigma\phi}\right] \tag{18}$$

The difference between Eqs. (17) and (18) lies in the fourth term on the right hand side. In Eq. (17), the fourth term is dependent on α, while in Eq. (18) it is not. These two relations have been largely used for the comparison of predicted barrier enhancement with either experimental or simulation results.

Minelli *et al.* [214] reported numerical and analytical modeling results for barrier properties in ordered nanocomposite systems. They proposed a new formulation capable of predicting the gas transport properties in simplified nanocomposite geometries and argued that their model correctly describes the enhancement in barrier effect for the systems for a wide range of filler loading and platelet dimensions, and can be reliably used to obtain relevant information on gas permeability in real nanocomposite systems. They proposed the following expression for the tortuosity factor:

$$\tau = \frac{\alpha\phi}{\sigma}\left(1+\frac{\sigma}{\alpha}\right)^2 + \frac{\alpha^2\phi^2(1+\sigma/\alpha)^4}{1-\phi(1+\sigma/\alpha)} + \frac{4\alpha\phi}{\pi}\left(1+\frac{\sigma}{\alpha}\right)^2 \ln\left[\frac{1-\phi(1+\sigma/\alpha)}{\sigma\phi(1+\sigma/\alpha)(\pi/2)}\right] \tag{19}$$

The first term on the right hand side of Eq. (19) results from the tortuous path contribution to the mass transport resistance and the remaining terms refer to the tortuous path for the diffusing molecule.

As we can see, several models have been developed to determine the barrier properties of nanocomposites. At this point, it is desirable to compare the prediction of different models presented herein. For this purpose, we consider three different models: Aris model [Eq. (17)], Wakeham and Mason model [Eq. (18)], and Minelli model [Eq. (19)]. Figure 15 demonstrates the outcome of this comparison, which are made considering the two pore aspect ratios (σ = 5 and 10); the value of α is taken as 800 since the aspect ratio of the graphene layer was roughly estimated to be about 800 by transmission electron microscope. [6] It can be observed that Wakeham and Mason model overestimates the barrier improvement in the dilute regime ($\phi \geq 0.002$) and all models predict almost the same results when the values of $\phi > 0.002$. Figure 15 also demonstrates that the values of σ marginally influence the barrier performance.

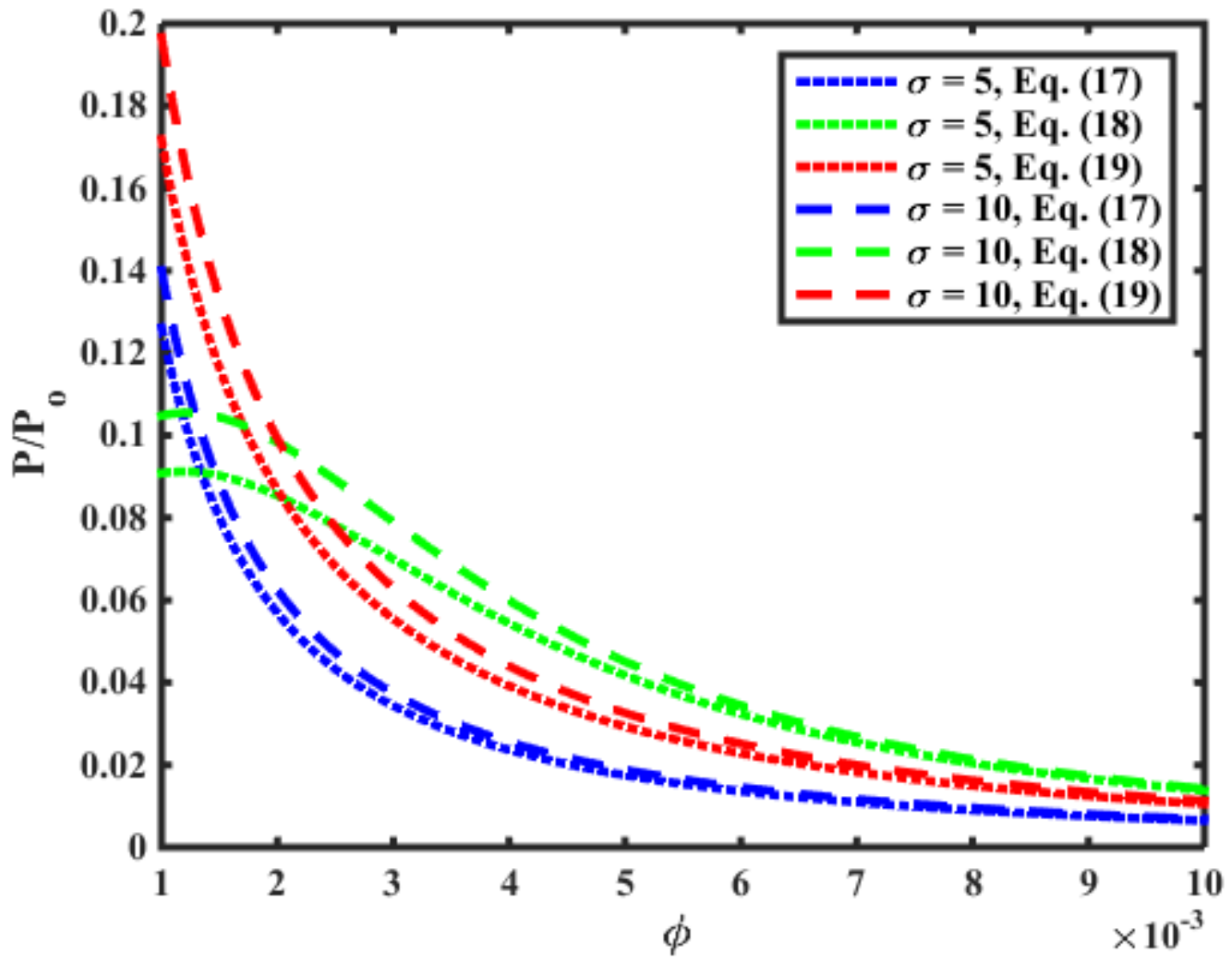

**Figure 15.** Comparison of theoretical model predictions ($\alpha = 800$).

In case of multilayer embedded graphene layers in the polymer matrix, we can use the following extended form of Aris' result: [212]

$$\tau = 1 + \frac{\alpha^2 \phi^2}{1 - \phi} + \frac{\alpha\phi}{\sigma} \tag{20}$$

Considering the effect of orientation of graphene layers, the modified Nielsen model can be utilized to determine the tortuosity factor: [11]

$$\tau = 1 + \frac{2\alpha\phi}{3}\left(S + \frac{1}{2}\right) \tag{21}$$

in which

$$S = \frac{1}{2}\langle 3\cos^2\theta - 1 \rangle, \quad -0.5 \leq S \leq 1 \tag{22}$$

where S is the order parameter representing the orientations of graphene layers, as demonstrated in Figure 16.

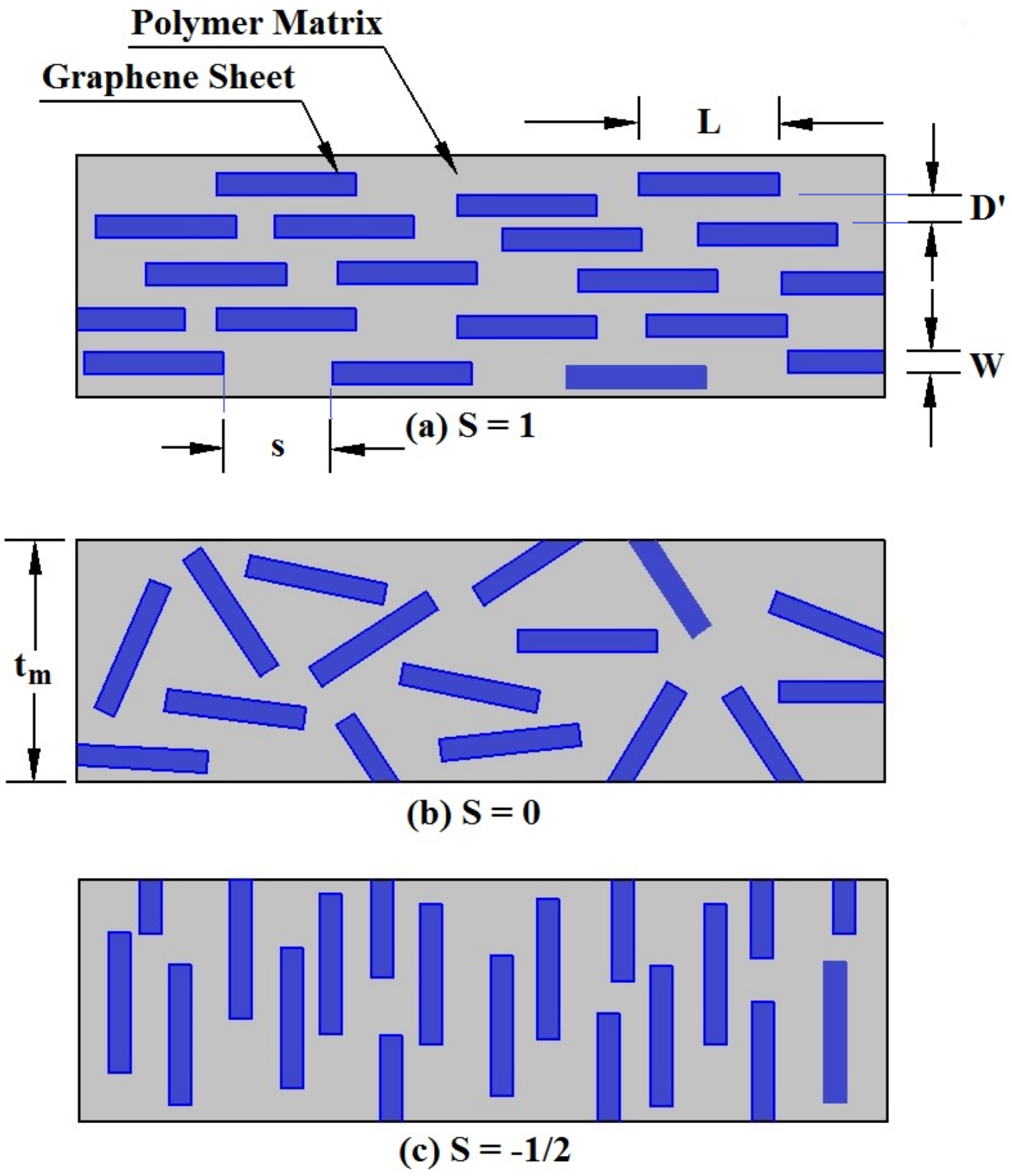

**Figure 16.** Different orientation of graphene layers in the nanocomposite film.

In case of high loading of graphene layers, Eq. (9) can be modified considering of the degree of staking with the parameter ⟨N⟩ as follows:

$$\tau = 1 + \frac{\alpha\phi}{\langle N\rangle} \tag{23}$$

Using Eqs. (22) and (23),

$$\tau = 1 + \frac{2\alpha\phi}{3\langle N\rangle}\left(S + \frac{1}{2}\right) \tag{24}$$

Referring to Figure 16, we can consider the three cases: perfect alignment of graphene layers, $S = 1$ ($\theta = 0^o$); (ii) random distribution of graphene layers, $S = 0$ ($\theta = 54.74^o$); and (iii) graphene layers do not provide barrier to the diffusion of gas molecules, $S = -1/2$ ($\theta = 90^o$). The effects of length, concentration and orientation of graphene layers, and degree of delamination on the relative permeability can be explored using Eq. (24). Dispersing longer graphene layers in a polymer matrix is particularly beneficial in several respects by (i) increasing the tortuosity, (ii) reducing the dependence of the relative permeability on the orientation order of the graphene layers, and (iii) slowing the degradation in barrier property with decreasing state of delamination, i.e., increasing aggregation via intercalation. The last of these factors ultimately controls the barrier properties of graphene/PNCs. The first two cases are considered herein to investigate their influence on the barrier performance of nanocomposites containing either a single or many layers of graphene. Figure 17 illustrates the variations of the relative permeability of nanocomposite with the volume fraction of graphene layers. It may be observed that the relative permeability of nanocomposite decreases when the graphene layers are horizontally dispersed ($S = 1$) in the matrix. For instance, in the case of graphene (N = 1) stacked horizontally in a polymer matrix with 1 vol% can reduce the gas permeation by ~89% in comparison to the pure polymer matrix. On the other hand, pronounced effect of random orientations ($S = 0$) of graphene layers on the barrier performance of the nanocomposite is observed. It may also be observed that the barrier performance of the nanocomposite decreases when the degree of graphene layer staking increases. This is due to the fact that the graphene layers tend to agglomerate with when the value of N increases and these agglomerates increase the gas permeation rate through the relatively highly permeable polymer matrix rather than graphene. Figure 17 clearly demonstrates that the barrier performance of multi-layer graphene/PNCs can be easily altered through effective utilization and orientation of graphene and its derivatives.

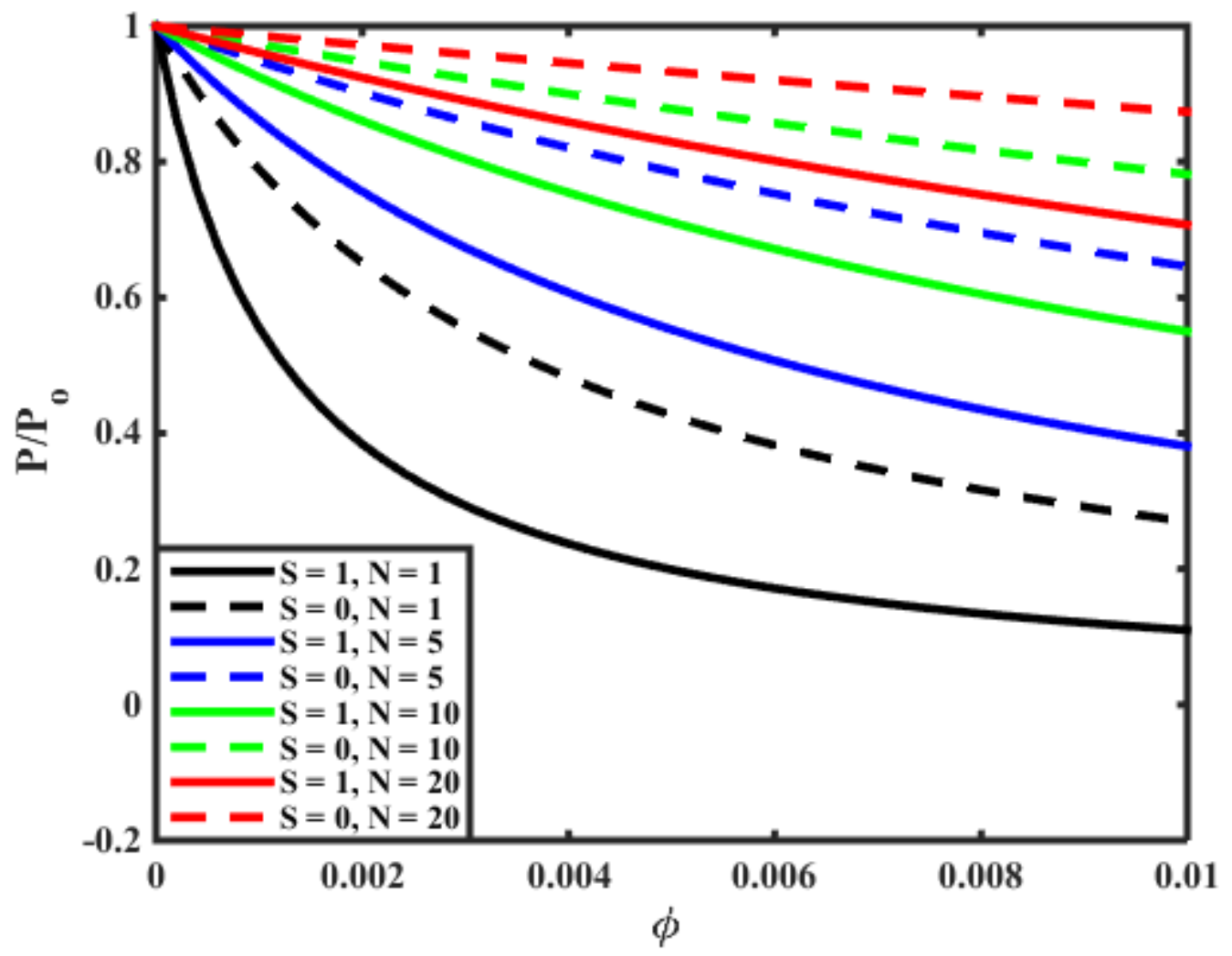

**Figure 17.** Effect of orientation and number of graphene layers on the barrier performance of nanocomposite ($\alpha = 800$).

Recently, Huang *et al.* [6] synthesized GONS/PLA films and reported their $O_2$ and $CO_2$ permeability coefficients. In their study, the enhanced gas barrier performance was ascribed to the remarkable impermeable property of GONSs as well as to the strong interfacial adhesion between GONSs and PLA matrix. Figure 18 illustrate the comparison between their experimental results and the analytical results predicted by Eq. (21). It may be observed that the experimental data are close to that of the analytical values, when S = 0, except the $O_2$ permeability coefficient at a GONS loading of 1.37 vol%. Such good agreement suggests that GONSs are prone to randomly disperse throughout the PLA matrix, and is in accordance with the GONSs morphology observed in their work. [6]

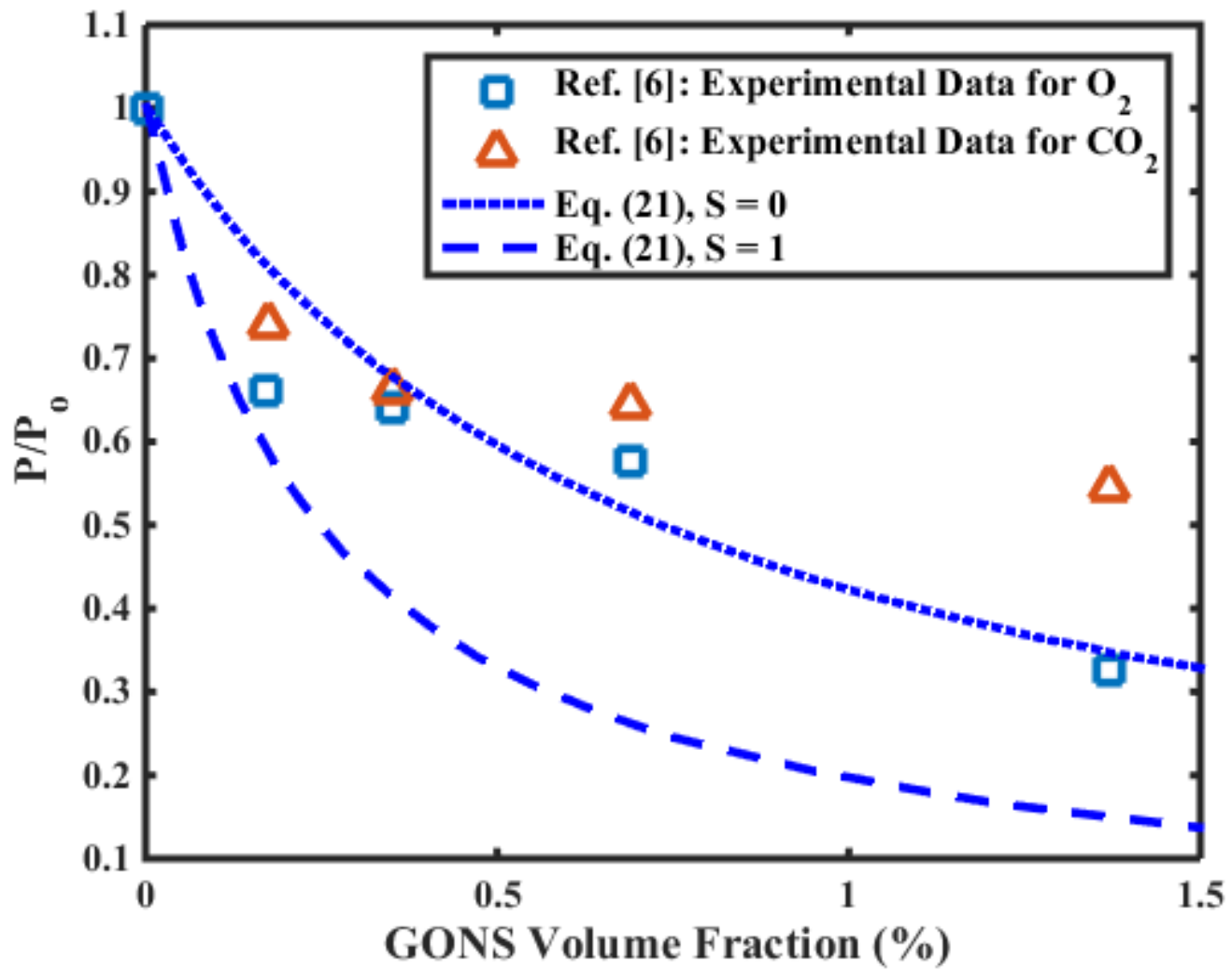

**Figure 18.** Comparison between the experimental data and the results predicted by Eq. (21) for the GONS/PLA films.

Sometimes nanocomposite may contain different sizes of graphene layers, and in such situation it is preferred to know which graphene size is dominating transport through a film. In this case, the tortuosity factor is given as follows: [203]

$$\tau = \left[1 + \left(\frac{\phi_{tot}}{3W\sum_i N_i l_i}\right)\sum_i N_i L_i^2\right]^2 \tag{25}$$

where $\phi_{tot}$ is the total volume fraction of graphene layers, $l_i$ is the average length of graphene layer in the size category i, and $N_i$ is the number of graphene layers in the size category i. It may be noted that Eq. (25) is derived assuming that graphene layers in adjacent layers do not influence each other. Therefore, this model is not suitable for high volume fractions of graphene layers. Picard *et al.* [199] modified this model accounting for the distribution of the filler thickness and contribution of the surfactant layer to the impermeable phase volume fraction for larger agglomerates. In case of graphene/PNCs, the modified tortuosity factor is given as follows:

$$\tau = \left[1 + \left(\frac{\phi_{tot}}{3\sum_i \frac{N_i l_i}{W_i}}\right)\sum_i \frac{N_i L_i^2}{W_i}\right]^2 \quad (26)$$

For the nanocomposite film containing random array of circular graphene layers, the tortuosity factor is given by: [202]

$$\tau = 4\left[\frac{1 + x + 0.1245x^2}{2 + x}\right]^2; \quad x = \frac{\pi D}{2W\ln\left(\frac{D}{2W}\right)}\phi \quad (27)$$

Similarly, for the nanocomposite film containing random dispersion of non-overlapping circular graphene layers, the tortuosity factor is: [215]

$$\tau = \exp\left[\left(\frac{D}{3.47W}\phi\right)^{0.71}\right] \quad (28)$$

where D denote the diameter of the graphene layer.

Accounting for the decrease in solubility, the modified form of original polynomial Eq. (27) can be used to determine the tortuosity factor for the graphene/PNCs film ; [204] as follows:

$$\tau = 1 - \phi + 4\left[\frac{1 + x + 0.1245x^2}{2 + x}\right]^2; \quad x = \frac{\pi D}{2W\ln\left(\frac{D}{2W}\right)}\phi \quad (29)$$

The Eq. (28) was proposed by Gusev and Lusti. [215] An exponential function of $(D/W)\phi$, was obtained by finite element calculations. Furthermore, this relation was modified by Picard *et al.* [199] with respect to the solubility effects; such that,

$$\tau = 1 + \frac{0.71D}{3.47W}\phi \quad (30)$$

## 5. Mixed matrix membranes for gas separation

Gas barrier performance of graphene/polymer nanocomposites has been studied so far in this work. However, mixed matrix membrane (MMM) comprising permeable/impermeable particles, such as zeolites, carbon molecular sieves (CMS), activated carbon, silica, metal organic framework and carbon nanotubes, and polymeric material presents an interesting approach for improving the separation properties of resulting membranes; the separation properties includes: permeability, selectivity, mechanical, thermal and chemical stability. Therefore, we herein discuss on the MMMs and present some of their permeability modeling techniques. In the recent years, gas separation membranes find many applications, such as carbon dioxide-nitrogen separation in flue gas, hydrogen-carbon dioxide separation for hydrogen production in fuel cells, oxygen–nitrogen separation for production of oxygen enriched air or pure nitrogen, carbon dioxide-methane separation for natural gas, vapour-vapour separation, and dehydration of air. [216, 217]

The inorganic materials used for MMMs can be classified into porous and nonporous types. [218] The effect of porous fillers on the MMM is different from nonporous inorganic fillers and can be related to their structure and pore size. Generally, porous fillers act as molecular sieving agents in the polymer matrix and separate gas molecules by their shape or size; [217, 219] these highly selective porous fillers allow the desired component to pass through the pores and thus the resulting MMM shows higher permselectivity compared to the neat polymeric membrane. Usually highly selective polymers improve the separation performance of MMMs, therefore, glassy polymers with superior gas selectivity are preferred to highly permeable but poorly selective rubbery polymers; [217, 220-224] but, the adhesion between the filler and surrounding glassy polymers is not perfect and weak organic-inorganic interaction between them causes voids at their interface. [225] In general the permeability of gas in the filler-polymer interfacial region is significantly different from the permeability of the bulk polymer and depends on the structure of this region on a nanoscale. As a consequence of the non-perfect filler-matrix interfacial contact: (i) the polymer chains in contact with the filler surface become rigidified in comparision to the bulk polymer, and (ii) de-wetting of polymer chains from the filler surface may occur resulting in the formation of voids in the interfacial region. [165] When the filler-matrix interfacial region becomes rigid then its permeability ($P_{int}$) is reduced by chain immobolzation factor, β, relative to that of the polymer; such that,

$$P_{int} = \frac{P_o}{\beta} \tag{31}$$

On the other hand, when the filler-polymer interfacial region conatins intarfacial voids, its permeability is larger than that of the polymer. Considering the pore size ($d_{pore}$) is larger than the size of the diffusive molecule ($d_{mol}$), the interfacial permeability ($P_{int}$) is given by the product of the Knudsen diffusivity ($D_{Knudsen}$) of a gas molecule

$$D_{Knudsen} = \frac{d_{pore}}{3}\left(1 - \frac{d_{mol}}{2W_{int}}\right)\sqrt{\frac{8RT}{\pi M}} \tag{32}$$

and the sorption coefficient ($S_{void}$) of the gas molecules in the void is given as

$$S_{void} = \frac{1}{RT}\left(1 - \frac{d_{mol}}{2W_{int}}\right)^2 \tag{33}$$

where R is the universal gas constant, T is the absolute temperature, and $W_{int}$ is the thickness of rigid filler-matrix interfacial region. In Eq. (32), the term $(1 - d_{mol}/2W_{int})$ is added to account for the finite size of the migrating molecule when the void dimension is of the same order as the gas molecule size.

Mixed matrix membrane permeation models can be classified to the following two models, one which is used to predict permeation of a MMM comprising permeable particles and polymer matrix and the other that is applicable for a MMM comprising impermeable particles and polymer matrix; accordingly, we introduce some important permeation models in this section to predict the barrier properties of MMMs.

### 5.1 MMMs containing permeable particles

Permeation models for MMMs containing permeable particles predict the effective permeability of a gaseous penetrant through the MMM as functions of its constituents and their volume fractions. Based on the shape factor of particles, Bouma *et al.* [226] modified Maxwell-Wagner-Siller model to predict the effective permeability of a MMM with a dilute dispersion of ellipsoids as follows:

$$P = P_o \frac{nP_p + (1-n)P_o - (1-n)\phi(P_o - P_p)}{nP_p + (1-n)P_o + n\phi(P_o - P_p)} \tag{34}$$

where n is the shape factor of particles and $P_p$ is the permeability of dispersed particles.

In Eq. (34), the limits of $n = 0$ and 1 lead to two-layer model and provide the following relations based on rules-of-mixture and inverse rules-of-mixture, respectively: [217, 219, 227]

$$P = \phi P_p + (1-\phi)P_o \tag{35}$$

$$\frac{1}{P} = \frac{\phi}{P_p} + \frac{(1-\phi)}{P_o} \tag{36}$$

Note that Eqs. (35) and (36) can be treated as series and parallel two-layer models, respectively. In case of random orientations of particles, we can utilize the geometric mean model: [228]

$$P = P_p^{\phi} + P_o^{(1-\phi)} \tag{37}$$

Eq. (34) reduces to the famous Maxwell's relation when the value of $n = 1/3$ corresponding to dilute suspension of spherical particles as follows: [219]

$$P = P_o \frac{P_P + 2P_o - 2\phi(P_o - P_P)}{P_P + 2P_o + 2\phi(P_o - P_P)} \tag{38}$$

Note that Eq. (38) holds true when the loading of spherical particles is less than 20% and interaction among adjacent particles is neglected. [219, 227, 228]It may also be noted that Eq. (38) does not account for particle size distribution, particle shape, and aggregation of particles: [217, 229]

For higher loadings of particles, we can utilize the following Bruggeman model: [230]

$$\left(\frac{P/P_o - P_p/P_o}{1 - P_p/P_o}\right)\left(\frac{P}{P_o}\right)^{-0.333} = 1 - \phi \tag{39}$$

Similar to Maxwell model, Eq. (39) also does not account for particle size distribution, particle shape, and aggregation of particles; to account such effects, the following Lewis-Nielsen model [229, 231, 232] can be used:

$$P = P_o \frac{1 + 2\phi\left(\frac{P_P/P_o - 1}{P_P/P_o + 2}\right)}{1 - \phi\psi\left(\frac{P_P/P_o - 1}{P_P/P_o + 2}\right)} \tag{40}$$

in which

$$\psi = 1 + \phi\left(\frac{1 - \phi_{max}}{\phi_{max}^2}\right) \tag{41}$$

where $\phi_{max}$ is the maximum packing volume fraction of particles

A Cussler model can be used to a dilute suspension of staggered flake spheres as follows:

$$P = P_o \frac{1}{1 - \phi + 1/\left(\frac{P_P}{\phi P_o} + 4\frac{1 - \phi}{\alpha^2\phi^2}\right)} \tag{42}$$

A generalized Maxwell model proposed by Petropoulous [233] and extended by Toy *et al.* [234] can be used to predict the permeation properties of MMM membrane containing randomly dispersed particles as follows:

$$P = P_o \left[ 1 + \frac{(1+G)\phi}{\left(\frac{P_P/P_o + G}{P_P/P_o - 1}\right) - \phi} \right] \tag{43}$$

where, G is a geometric factor accounting for the effect of dispersed particle's shape: G = 1, for long and cylindrical particles which are dispersed transverse to the gas flow direction; G = 2, for spherical particles or isometric aggregates; G = ∞, for planar particles which are oriented parallel direction to the gas flow direction; G = 0, for dispersed particles which are oriented in perpendicular direction to the gas flow direction.

### 5.2 MMMs containing impermeable particles

Impermeable particles improve the gas separation properties of the resulting MMM membranes by increasing the matrix tortuous pattern and decreasing the diffusion of the larger molecules. [217, 235] Nano-scale inorganic particles may also disrupt the polymer chain packing and increase the free volume between polymer chains, which eventually increases the gas diffusion across a MMM membrane. Furthermore, the hydroxyl and other functional groups on the surface of these nanomaterials may also interact with polar gases ($CO_2$ and $SO_2$) and thus improve the penetrant solubility in the resultant MMM membranes. [217, 236] The developed permeation models in section 4.1 for grapheme nanocomposites can be used straight forward in the case of MMMs containing impermeable particles.

### 5.3 Permeation models for three-phase MMMs

So far, in this study, the permeation models for two-phase MMMs are shown considering perfect contact between a particle and the surrounding polymer matrix. In actual practice, particles are not perfectly bonded to the polymer matrix and poor adhesion occurs between them; such non-perfect bonding situation leads to (i) formation of a rigidified polymer layer around particles, (ii) pore blockage in porous particles, and (iii) creation of voids at the particle-polymer matrix interface. The existence of an interfacial layer between a particle and the surrounding polymer matric can be considered as a third phase of resulting MMM and three-phase permeation models can be used, as follows:

(i) Maxwell's modified model: [219, 229]

$$P = P_o \frac{2(1-\phi) + (1+2\phi)(P_{eff}/P_o)}{(2+\phi) + (1-\phi)(P_{eff}/P_o)} \tag{44}$$

in which

$$P_{eff} = P_i \frac{2(1-\phi_s) + (1+2\phi_s)(P_p/P_i)}{(2+\phi_s) + (1-\phi_s)(P_p/P_i)}$$

(ii) Felske model: [229, 237]

$$P = P_o \frac{2(1-\phi) + (1+2\phi)(\beta/\gamma)}{(2+\phi) + (1-\phi)(\beta/\gamma)} \tag{45}$$

(iii) Felske modified model: [238-240]

$$P = P_o \frac{1 + 2\phi(\beta-\gamma)/(\beta+2\gamma)}{1 - 2\psi(\beta-\gamma)/(\beta+2\gamma)} \tag{46}$$

where $P_{eff}$ is the permeability of two-phase particle-interphase medium, $P_i$ is the permeability of the interphase, $\phi_s$ is the volume fraction of an inorganic phase in the combined inorganic and interphase, $\gamma$ is the ratio of an interphase to the particle radius, and $\psi$ parameter described as a function of loading of particles.

## 6. Conclusion

Graphene and its derivatives have been identified as a powerful candidate for gas-barrier materials because perfect graphene do not allow diffusion of small gases through its plane. Graphene-incorporated polymers can not only enhance gas-barrier properties but also enhance mechanical strength and improve electrical conductivity and thermal properties when properly dispersed in a polymer matrix. In this paper, a thorough review of the production methods and gas barrier properties of PNCs composed of various polymer matrices and graphene have been presented. Many approaches have been attempted to prepare graphene/PNCs, including melt and solution mixing, in situ polymerization, layer-by-layer and latex coagulation, etc. For graphene/PNCs, the size, stacking orientation and exfoliation of graphene nanosheets in the polymer matrix are the most important factors that govern the gas barrier properties of nanocomposites. Compared to pure polymer matrix, the gas barrier performance of graphene/PNCs are often improved by 1-3 orders. To achieve superior gas barrier performance graphene/PNCs, the structure-property relationships of graphene/polymer also need to be fully understood. Further improvements in the gas barrier properties of graphene/polymer nanocmoposites could be expected from the development of more compatible graphene/polymer systems, complete exfoliation and homogeneous dispersion of graphene layers in polymer matrix and methods to prevent aggregation of graphene nanosheets in a polymer matrix and enhance their structural stability at high temperature. Lastly, the mathematical modeling aspects of gas barrier properties of graphene/PNCs are also thoroughly addressed. These diffusion models are useful to evaluate the gas barrier performance the PNCs before actually making them and identify ideal set of geometric and material parameters of the nanofillers, and their optimal loading in a given matrix that lead to optimum gas barrier performance of PNCs.

## Acknowledgments

Dr. S. Kumar and Dr. Yanbin Cui gratefully acknowledge the financial support from Borouge Pet. Ltd. (Project No: EX2014-000027).